\documentclass{amsart}

\usepackage[margin=1in]{geometry}
\usepackage{amsmath, amssymb, amsthm}
\usepackage{xcolor}
\usepackage{multirow}
\usepackage{subfigure}

\usepackage[lined,boxed,commentsnumbered,ruled]{algorithm2e}

\usepackage{verbatim}  
 
\numberwithin{equation}{section}  

\usepackage{algorithmic}

\usepackage{graphicx, epstopdf}

\usepackage{bm}

\newcommand{\image}{\textbf{\rm i}}

\newcommand{\bx}{\bm{x}}
\newcommand{\bv}{\bm{v}}
\newcommand{\bu}{\bm{u}}

\newcommand{\bq}{\bm{q}}

\newcommand{\bk}{\bm{k}}
\newcommand{\bp}{\bm{p}}
\newcommand{\hf}{\hat{f}}

\newcommand{\bZ}{\mathbb{Z}}
\newcommand{\bR}{\mathbb{R}}
\newcommand{\bS}{\mathbb{S}}

\newcommand{\mF}{\mathcal{F}}
\newcommand{\mD}{\mathcal{D}}
\newcommand{\mQ}{\mathcal{Q}}

\newcommand{\hB}{\hat{B}}
\newcommand{\mB}{\mathcal{B}}

\newcommand{\dd}{{\rm d}}
\newcommand{\Kn}{\mathit{Kn}}
\newcommand{\Ma}{\mathit{Ma}}

\title[Steady-state solver for the Boltzmann equation]{A symmetric Gauss-Seidel method for the steady-state Boltzmann equation}

\author{Tianai Yin}
\address[Tianai Yin]{
Beijing Computational Science Research Center, Beijing, China 100193}
\email{tianai.yin@csrc.ac.cn}

\author{Zhenning Cai}
\address[Zhenning Cai]{Department of Mathematics, National University of Singapore,
  Level 4, Block S17, 10 Lower Kent Ridge Road, Singapore 119076}
\email{matcz@nus.edu.sg}

\author{Yanli Wang}
\address[Yanli Wang]{Beijing
    Computational Science Research Center, Beijing, China 100193}
\email{ylwang@csrc.ac.cn}

\thanks{Tianai Yin was supported by the China Scholarship Council (CSC) (File No.202204890003). Zhenning Cai's work was supported by the Academic Research Fund of the Ministry of Education of Singapore under grant number A-8000965-00-00. This work of Yanli Wang is partially supported by the National Natural Science Foundation of China (Grant No. 12171026, U2230402, and 12031013), and the Foundation of the President of China Academy of Engineering Physics (YZJJZQ2022017). }

\begin{document} 
\maketitle

\begin{abstract}
We introduce numerical solvers for the steady-state Boltzmann equation based on the symmetric Gauss-Seidel (SGS) method. 
Due to the quadratic collision operator in the Boltzmann equation, the SGS method requires solving a nonlinear system on each grid cell, and we consider two methods, namely Newton's method and the fixed-point iteration, in our numerical tests.
For small Knudsen numbers, our method has an efficiency between the classical source iteration and the modern generalized synthetic iterative scheme, and the complexity of its implementation is closer to the source iteration.
A variety of numerical tests are carried out to demonstrate its performance, and it is concluded that the proposed method is suitable for applications with moderate to large Knudsen numbers.

\end{abstract}

\section{Introduction} \label{sec:introduction}
Rarefied gas dynamics is a branch of fluid dynamics that arises when the density of the gas is low (or when the characteristic length is tiny) and the discontinuous particle effect becomes more pronounced. It plays a vital role in astronautics and micro/nano flows \cite{shen2006rarefied}.
In astronautics, rarefied gas dynamics finds applications in many aerospace fields, including missiles, spacecraft, space shuttles, and space stations. It is also essential in micro-electro-mechanical systems, where the devices can be as tiny as a few microns. In these applications, rarefaction effects such as the gas-wall friction and the cold-to-hot heat transfer may dominate the gas flows. Simulations using rarefied gas models are therefore needed in such circumstances.

To describe how rarefied the gas is, people usually use a dimensionless parameter known to be the Knudsen number $\Kn$, which is the ratio of the mean free path of gas molecules to the characteristic length of the problem. It is generally believed that the continuum hypothesis, which is the basis of the Euler equations and Navier-Stokes equations, fails to hold when the Knudsen number is greater than $0.1$. In this regime, the gas kinetic theory needs to be adopted for accurate simulations, and one of the fundamental models is the Boltzmann equation, which describes the evolution of the distribution function of gas molecules.

Solving the Boltzmann equation numerically has been a challenging task even with modern supercomputers, owing to its high dimensionality, complicated collision terms, and the gas-surface interaction that may ruin the regularity of the solution. At present, the mainstream numerical methods mainly fall into two categories: direct simulation of Monte Carlo (DSMC) and deterministic methods. The DSMC method \cite{bird1964,MolecularBird} is a stochastic approach using simulation particles to model a cluster of gas molecules. It is a powerful and efficient solver for highly rarefied gas flows, but in the region where the gas is dense (e.g. $\Kn < 0.1$), its efficiency is weakened and the statistical noise may become strong. In recent decades, many deterministic methods have been developed to achieve noise-free solutions. Regarding the discretization of the collision term in the Boltzmann equation, the classical discrete velocity method \cite{DVM1964Broadwell,DVM1980,DVM1989} has evolved into spectral methods with remarkably higher efficiency, and spectral methods can also be categorized into two types: schemes on unbounded domains and schemes on truncated domains. The first category is a natural choice since the velocity domain is unbounded in the original Boltzmann equation, and different choices of basis functions lead to various methods, including the Hermite spectral method \cite{Sarna2018,Kitzler2019,Hu2020}, the Burnett spectral method \cite{Gamba2018,Hu2020Burnett,CaiBurnett2020,CaiBurnett2022Adaptive}, and the mapped Chebyshev method \cite{Hu2020Petrov,Hu2022}. However, these methods may suffer from large computational costs when discretizing the binary collision term. The second class of methods that apply the Fourier spectral method on truncated and periodized domains has a significantly lower computational cost. While the original version \cite{Pareschi1996} has the computational cost $O(N^6)$ ($N$ is the number of modes in each direction), more efficient schemes with the computational complexity $O(N^{\alpha} \log N)$, $\alpha \leqslant 4$ have been developed for a general class of collision models \cite{mouhot2006fast,Wu2013, Hu2017}. Due to the truncation of the domain, some properties such as momentum and energy conservation are lost during the discretization, which needs to be fixed after every time step by a post-processing \cite{Gamba2009, Cai2024}.

Efficient spatial discretization of the Boltzmann equation has also been widely studied. Examples include the fast kinetic scheme \cite{FKS_2018}, which focuses on the discretization of the convection term, and the series of unified gas kinetic schemes (UGKS) introduced by Xu et al. in their work \cite{XU2010}, which takes into account the integrated effect of transport and collision in the numerical scheme.
The UGKS combines the kinetic theory and methods in computational fluid dynamics, and can be regarded as a multiscale method based on the direct modeling of flow physics. 
The method's applicability has been extended to simulations involving plasma \cite{liu_xu_2017} and dilute gas \cite{Liu2019}. Nevertheless, when dealing with intricately structured flows, integral solutions may fall short of providing adequately accurate time evolution unless more intricate numerical modeling of local physics is employed, especially on coarse grids and with large time steps, as indicated in a recent study by Zhu et al. \cite{Zhu2021decade}. More recently, an innovative particle-based approach, known as the Unified Gas Kinetic Wave-Particle Method (UGKWP), was proposed in \cite{LIU2020108977}, building upon the UGKS foundation. The UGKWP method leverages stochastic particle simulations in conjunction with the original UGKS to enhance computational efficiency, particularly for hypersonic flow simulations across all regimes. For a comprehensive understanding, additional details can be found in \cite{Zhu2019} and the associated references.
Another popular approach for both spatial and velocity discretizations is the dynamical low-rank method \cite{lowrank2019,lowrank2021}, which assumes that the spatial and velocity variables are weakly entangled, so that the dimensionality of the solution space is considerably reduced.

In this work, we focus mainly on iterative methods of the steady-state nonlinear Boltzmann equation. A classical iterative scheme, known as the source iteration, treats the kinetic equation as a fixed-point problem, so that the fixed-point iteration can be applied \cite{Adams2002}. However, this method suffers from slow convergence rates when the interaction between particles is strong. To accelerate the iteration, the synthetic method, which uses equations in the asymptotic limit as a preconditioner, was proposed in \cite{Kopp1963}. Recently, such an approach has been further developed into the general synthetic iterative scheme (GSIS) \cite{Su2020}, which can be applied to different types of equations and is shown to be efficient in all regimes \cite{Wu2020fast}. The application to the nonlinear Boltzmann equation can be found in \cite{Zhu2021}. Using the Navier-Stokes equation as a preconditioner, the method eliminates the problem of slow convergence for small Knudsen numbers, and compared with the source iteration, the number of iterations is reduced for almost all test cases. To achieve these benefits, one extra part in the implementation is to solve the steady-state synthetic equations (which are essentially the Navier-Stokes equations), which might sometimes be challenging when the domain or the fluid structure is complicated. The aim of our work is to investigate numerical schemes without solving the macroscopic equations, whose implementation is relatively easier. Compared with GSIS, our method will require more computational cost for smaller Knudsen numbers, but the efficiency will still be significantly higher than the source iteration.

In general, our method will be based on the Fourier spectral method for the velocity discretization and the symmetric Gauss-Seidel iterations over the spatial grid. When updating the solution on each spatial grid cell, both Newton's method and the fixed-point iteration are tested. Our methods will be introduced after a review of the Boltzmann equation and the Fourier spectral method in Section \ref{sec:preliminaries}. Both numerical schemes, as well as some analysis of the convergence rates, are detailed in Sec. \ref{sec:numerical_scheme}. Our numerical results are demonstrated in Sec. \ref{sec:numerical_experiments}, and we provide some concluding remarks in Sec. \ref{sec:conclusion}.

\section{Preliminaries} \label{sec:preliminaries}

\subsection{Boltzmann equation}
The Boltzmann equation is devised based on the kinetic description of gases by Ludwig Boltzmann in 1872, in which the fluid state is described by a probability distribution function (phase density function) $f(t,\bx,\bv)$, where $t$ is the time variable, and $\bx$ and $\bv$ represent the spatial location and velocity of the gas molecules, respectively. For a single-species, monatomic gas without external forces, the Boltzmann equation has the following form
\begin{equation}
    \frac{\partial f(t,\bx,\bv)}{\partial t}
    +\bv\cdot\nabla_{\bx} f(t,\bx,\bv)
    =\mQ[f,f](t,\bx,\bv),
    \qquad t \in \mathbb{R}^+, \quad \bx \in \Omega, \quad \bv \in \mathbb{R}^3.
    \label{eq:Boltzmann_origin}
\end{equation}
The right-hand term $\mQ[f,f]$ describes binary collisions between gas molecules \cite{MolecularBird} and can be written as
\begin{equation}
    \mQ[f,f](t,\bx,\bv)
    =\int_{\bR^3} \int_{\bS^2}
    \mB(\bv-\bv_{\ast},\sigma)[f(t,\bx,\bv')f(t,\bx,\bv'_{\ast})-f(t,\bx,\bv)f(t,\bx,\bv_{\ast})]
    \dd \sigma\dd \bv_{\ast},
    \label{eq:Q_origin}
\end{equation}
where the pre-collision velocities are $\bv$ and $\bv_{\ast}$, from which the post-collision velocities can be obtained by
\begin{equation}
    \bv' = \frac{\bv+\bv_{\ast}}{2}+\frac{|\bv-\bv_{\ast}|}{2}\sigma, 
    \qquad \bv_{\ast}'=\frac{\bv+\bv_{\ast}}{2}-\frac{|\bv-\bv_{\ast}|}{2}\sigma.
\label{eq:postcollision}
\end{equation}
The nonnegative collision kernel $\mB$ has the form
\begin{equation}
    \mB(\bv-\bv_{\ast},\sigma)=b(|\bv-\bv_{\ast}|,\cos \vartheta)=|\bv-\bv_{\ast}|\cdot \frac{b}{\sin \vartheta}\left|\frac{\dd b}{\dd \vartheta}\right|,\qquad
    \cos \vartheta=\frac{\sigma \cdot(\bv-\bv_{\ast})}{|\bv-\bv_{\ast}|}.
    \label{eq:mB}
\end{equation}
In the above formula, the symbol $b$ stands for the miss-distance impact parameter, and the deflection angle $\vartheta$ in 
the variable hard sphere (VHS) model and
the variable soft sphere (VSS) model
are given by   
\begin{equation}
\vartheta=2\cos^{-1}(b/d)\quad {\rm (VHS)},\qquad
\vartheta=2\cos^{-1}((b/d)^{1/{\alpha}}) \quad {\rm (VSS)},
\label{eq:chi_VHS}
\end{equation}
where $\alpha$ is the scatting parameter, and $d$ is the diameter of gas molecules.
In both models, the diameter $d$ is a function of the relative speed $|\bv-\bv_{\ast}|$. More details regarding the VHS models are given in Appendix \ref{sec:non-dimensionalization}.

In our work, we focus on the steady-state Boltzmann equation which removes the time derivative from \eqref{eq:Boltzmann_origin}.
For simplicity, we study its dimensionless form given by
\begin{equation}
\bv\cdot\nabla_{\bx}f=\frac{1}{Kn}\mQ[f,f], \qquad \bx \in \Omega, \quad \bv \in \mathbb{R}^3,
    \label{eq:Boltzmann_steady}
\end{equation}
where $\Kn$ is the Knudsen number as introduced in Sec. \ref{sec:introduction}.
The details of the nondimensionalization can be found in Appendix \ref{sec:non-dimensionalization}.
To determine the solution, boundary conditions need to be specified on $\partial \Omega$. 
The general form of boundary conditions is
\begin{displaymath}
f(\bx,\bv) = g(\bx,\bv), \qquad \text{for } \bx \in \partial\Omega \text{ and } \bv \cdot \boldsymbol{n}(\bx) < 0,
\end{displaymath}
where $\boldsymbol{n}(\bx)$ is the outer unit normal vector at the boundary point $\bx$.
In this work, two types of boundary conditions, namely the inflow boundary condition and the wall boundary condition, will be studied, and the details will be given in Sec. \ref{sec:bc}.
Note that when wall boundary conditions are specified on the entire boundary $\partial \Omega$, an additional condition specifying the total mass
\begin{equation} \label{eq:total_mass}
\int_{\Omega} \int_{\mathbb{R}^3} f(\bx,\bv) \,\mathrm{d}\bv \,\mathrm{d}\bx = M
\end{equation}
is needed to uniquely determine the solution.

In most cases, instead of the distribution function itself, people are more interested in its moments, including the density $\rho$, the macroscopic velocity $\bu$ and the thermal temperature $T$:
\begin{equation}
    \rho=\int_{\bR^3} f\dd \bv,
    \qquad
    \bu=\frac{1}{\rho}\int_{\bR^3} \bv f\dd \bv,
    \qquad
    T=\frac{1}{3\rho}\int_{\bR^3} |\bv-\bu|^2 f\dd \bv,
    \label{eq:rho_u_T}
\end{equation}
as well as the heat flux $\bq$ and the pressure tensor $p_{ij}$ defined as
\begin{equation}
    \bq=\frac{1}{2}\int_{\bR^3}
    |\bv-\bu|^2(\bv-\bu)f\dd\bv,
    \qquad
    p_{ij}=\int_{\bR^3}(v_i-u_i)(v_j-u_j)f\dd \bv,\qquad i,j=1,2,3.
    \label{eq:qi_pij}
\end{equation}

\subsection{Fourier spectral method} \label{sec:Fourier}
Due to the complicated form and the high-dimensionality of the binary collision term \eqref{eq:Q_origin}, the most costly part of the Boltzmann numerical solver is the collision term. Currently, the Fourier spectral method is one of the most popular and efficient deterministic methods in the discretization of the collision term. Our work will also be based on this approach.

Below, we will summarize the Fourier spectral method for computing the binary collision term $\mathcal{Q}$. Following \cite{Pareschi1996}, 
the Fourier spectral method in the velocity space is derived by assuming that $f(\bv)$ has a compact support: $\operatorname{supp} (f) \subset B_R := \{\bv \in \mathbb{R}^3 \mid |\bv| < R\}$.
This assumption can be approximately satisfied since $f \in L^1(\mathbb{R}^d)$.
One can then truncate the domain of $\bv$ to $\mD_{L_v}=[-L_v, L_v]^d$ with $L_v \geqslant \frac{3+\sqrt{2}}{2} R$ and periodize both the function and the collision operator. For more details, we refer the readers to \cite{dimarco_pareschi_2014}
and here we only provide the final form: assuming that $f(\bv)$ has the Fourier expansion
\begin{displaymath}
f(\bv) = \sum_{\bp \in \mathbb{Z}^d} \hat{f}_{\bp} E_{\bp}(\bv), \qquad E_{\bp} (\bv) = \exp\left(\frac{\image \pi}{L_v} \bp \cdot \bv\right),
\end{displaymath}
we can expand the truncated and periodized collision operator $\mathcal{Q}^R$ into the Fourier basis:
\begin{equation} \label{eq:QRff}
    \mQ^R[f,f](\bv)=
    \sum_{\boldsymbol{l}\in \bZ^d} \sum_{\boldsymbol{m}\in \bZ^d} 
    (\tilde{B}(\boldsymbol{l},\boldsymbol{m})-\tilde{B}(\boldsymbol{m},\boldsymbol{m})) \hat{f}_{\boldsymbol{l}}\hat{f}_{\boldsymbol{m}} 
    E_{\boldsymbol{l}+\boldsymbol{m}}(v),
\end{equation}
where $\tilde{B}(\boldsymbol{l}, \boldsymbol{m})$ is related to the collision kernel by
\begin{displaymath}
\tilde{B}(\boldsymbol{l}, \boldsymbol{m}) = \int_{B_R} \int_{\mathbb{S}^2} \mathcal{B}(\boldsymbol{g}, \boldsymbol{\sigma}) E_{\boldsymbol{l}} \left( \frac{\boldsymbol{g} + |\boldsymbol{g}| \boldsymbol{\sigma}}{2}  \right) E_{\boldsymbol{m}} \left( \frac{\boldsymbol{g} - |\boldsymbol{g}| \boldsymbol{\sigma}}{2}  \right) \,\mathrm{d}\boldsymbol{\sigma}\,\mathrm{d}\boldsymbol{g}.
\label{eq:tilde_B}
\end{displaymath}
For simplicity, below we use the notation $\mathcal{F}_{\bp}(f)$ to denote the $\bp$-th Fourier coefficient of $f(\bv)$:
\begin{displaymath}
    \hf_{\bp}=
    \mF_{\bp}(f)
    =\frac{1}{(2L_v)^d}
    \int_{\mD_{L_v}}f(\bv)E_{-\bp}(\bv) \dd \bv.
\end{displaymath}
Then the $\bk$-th Fourier coefficient
of
equation \eqref{eq:QRff} can be written as
\begin{equation} \label{eq:fsm}
    \mathcal{F}_{\boldsymbol{k}}(Q^R[f,f]) = \sum_{\boldsymbol{l},\boldsymbol{m}\in\bZ^d}\mathbf{1}(\boldsymbol{l}+\boldsymbol{m}-\boldsymbol{k})
     (\tilde{B}(\boldsymbol{l},\boldsymbol{m})-\tilde{B}(\boldsymbol{m},\boldsymbol{m})) \hat{f}_{\boldsymbol{l}}\hat{f}_{\boldsymbol{m}}.
\end{equation}
The Fourier spectral method is a direct truncation of $\mathbb{Z}^d$ to a finite set $\{-n,\cdots,n\}^d$. For simplicity, below we will use $N  = 2n+1$ to denote the number of modes in each direction.

\subsection{Source iteration} \label{sec:source_iter}
A classical method for the steady-state Boltzmann equation \eqref{eq:Boltzmann_steady} is the source iteration based on the following iterative method:
\begin{equation} \label{eq:source_iter}
\bv \cdot \nabla_{\bx} f^{n+1} + \frac{\nu}{\Kn} f^{n+1} = \frac{1}{\Kn} (\mathcal{Q}[f^n,f^n] + \nu f^{n}),
\end{equation}
where $f^n$ refers to the solution of the $n$-th iteration. However, it is well-known that such an iteration converges slowly when $\Kn$ is small \cite{Adams2002, SU_GSIS2020}. Briefly speaking, the reason is that when $\Kn$ is small, the distribution functions are close to local Maxwellians, so that $Q^R[f^n, f^n] \sim O(\Kn)$ is close to zero. Thus, the scheme \eqref{eq:source_iter} has the form
\begin{displaymath}
f^{n+1} = f^n + O(\Kn),
\end{displaymath}
meaning that $f^{n+1}$ is only a tiny update of $f^n$, so that the scheme converges slowly.

In this work, we will test a different approach to solving the steady-state Boltzmann equation, which will be detailed in the next section. For simplicity, we restrict ourselves to the 1D3V case, meaning one spatial dimension and three velocity dimensions, so that the steady-state Boltzmann equation has the form
\begin{displaymath}
v_1 \partial_x f = \frac{1}{\Kn} \mathcal{Q}[f,f],
\end{displaymath}
where $v_1$ is the first component of the velocity vector $\bv$.

\section{Numerical scheme} \label{sec:numerical_scheme}
In this section, we introduce two numerical schemes to solve the 1D3V steady-state equation. The Fourier spectral method is applied to the collision term, which requires us to truncate the velocity domain to $[-L_v, L_v]^3$ as mentioned in Sec. \ref{sec:Fourier}. The discrete velocity method defined on the collocation points is applied to the advection term, and the conversion can be done by the fast Fourier transform. The spatial domain is assumed to be $\Omega = 
[x_{\rm L},x_{\rm R}]\subset \bR$, which is discretized with a uniform grid with $N_x$ cells, so that the size of each cell is $\Delta x = (x_{\rm R} - x_{\rm L}) / N_x$.  

Below we use $f_{\bk,j}$ with $j=1,\ldots,Nx$ to denote the numerical approximation to the average of the distribution function $f(\cdot,\bv_{\bk})$ in the $j$-th grid cell.
Based on the discrete velocity method, we use the upwind scheme with linear reconstruction for the transport term to obtain the fully discrete scheme: 

\begin{equation}
    \frac{v_{1,\bk}^+}{\Delta  x}\left(f_{\bk,j+1/2}^{L} -
    f_{\bk,j-1/2}^{L}\right)
    + 
    \frac{v_{1,\bk}^-}{\Delta  x}\left(f_{\bk,j+1/2}^{R} 
    - f_{\bk,j-1/2}^{R}\right)
    =\frac{1}{\Kn}Q_{\bk}(\boldsymbol{f}_j),
    \label{eq:reconstruction_equation}
\end{equation}
where $\boldsymbol{f}_j$ is the vector including $f_{\bk,j}$ for all $\bk$, and $Q_{\bk}(\boldsymbol{f}_j)$ 
represents the approximation of the collision term 
$\mQ[f,f]$ at the point $(x_j, \bv_{\bk})$. The terms $f_{\bk,j+1/2}^{L,R}$ are obtained from the linear reconstruction:
\begin{equation} \label{eq:reconstruction}
      f_{\bk,j+1/2}^{R}=f_{\bk,j+1}-s_{\bk,j+1}\frac{\Delta  x}{2},
      \qquad 
      f_{\bk,j+1/2}^{L}=f_{\bk,j}+s_{\bk,j}\frac{\Delta  x}{2},
\end{equation}
and the slope $s_{\bk,j}$ in grid point $(x_j,\bv_{\bk})$ is obtained as follows:
\begin{equation}
s_{\bk,j}
 = \left\{
    \begin{aligned}
      &\frac{f_{\bk,1}-f_{\bk,0}}{\Delta x},&j=1;  \\
     &\frac{f_{\bk,j+1}-f_{\bk,j-1}}{2\Delta x},&1<j<N_x ; \\
      &\frac{f_{\bk,N_x-1}-f_{\bk,N_x-2}}{\Delta x},& j=N_x.
     \end{aligned}\right.
    \label{eq:slope}
\end{equation}
In \eqref{eq:reconstruction_equation}, the right-hand side is defined by the Fourier spectral method applied to the interpolated distribution function:
\begin{displaymath}
  Q_{\bk}(\boldsymbol{f}_j) = \mathcal{P}_N \mathcal{Q}^R[\mathcal{I}_N(\boldsymbol{f}_j),\mathcal{I}_N(\boldsymbol{f}_j)] \Big|_{\bv = \bv_{\bk}}.
\end{displaymath}
Here $\mathcal{I}_N$ and $\mathcal{P}_N$ are, respectively, the interpolation and projection operators of the Fourier spectral method. To utilize the fast solver \eqref{eq:fsm} in the computation, fast Fourier transforms are applied to implement the transformation between the physical and the frequency spaces.

To solve the nonlinear system \eqref{eq:reconstruction_equation}, we adopt the symmetric Gauss-Seidel (SGS) iteration in our solver. To demonstrate the idea, we use the first-order discretization ($s_{\bk,j} = 0$) as an example. Let $n$ be the index of the iteration. Our scheme reads
\begin{align}
    \label{eq:SGS1}
    & \frac{v_{1,\bk}^+}{\Delta  x}\left(f_{\bk,j}^{n*} -
    f_{\bk,j-1}^{n*}\right)
    + 
    \frac{v_{1,\bk}^-}{\Delta  x}\left(f_{\bk,j+1}^n
    - f_{\bk,j}^{n*}\right)
    =\frac{1}{\Kn}Q_{\bk}(\boldsymbol{f}_j^{n*}), \\
    \label{eq:SGS2}
    & \frac{v_{1,\bk}^+}{\Delta  x}\left(f_{\bk,j}^{n+1} -
    f_{\bk,j-1}^{n*}\right)
    + 
    \frac{v_{1,\bk}^-}{\Delta  x}\left(f_{\bk,j+1}^{n+1} 
    - f_{\bk,j}^{n+1}\right)
    =\frac{1}{\Kn}Q_{\bk}(\boldsymbol{f}_j^{n+1}).
\end{align}
Similar to the SGS method, the grid cells are scanned twice in each step, and during each scan, we need to solve a nonlinear system to obtain the distribution function in each grid cell. The nonlinear solver will be introduced in the following subsections. Here we just mention that in the second-order scheme, a correction term in the form $s_{\bk,j} \Delta x / 2$ is added to each $f$ on the left-hand sides of \eqref{eq:SGS1}\eqref{eq:SGS2} (see \eqref{eq:reconstruction}), and here the slope $s_{\bk,j}$ is always defined by the solution the $n$-th iteration solution $f^n$, meaning that $s_{\bk,j}^n$ is always used in \eqref{eq:SGS1} and $s_{\bk,j}^{n\ast}=s_{\bk,j}^n$ is used in \eqref{eq:SGS2}. 

Compared with the source iteration, the convergence of the SGS iteration does not stagnate when $\Kn$ approaches zero. When $\Kn$ is small, the equations \eqref{eq:SGS1} and \eqref{eq:SGS2} indicate that
\begin{displaymath}
\boldsymbol{f}_j^{n*} = \boldsymbol{\mathcal{M}}_j^{n*} + O(\Kn), \qquad
\boldsymbol{f}_j^{n} = \boldsymbol{\mathcal{M}}_j^{n} + O(\Kn),
\end{displaymath}
where $\boldsymbol{\mathcal{M}}_j^{n*}$ and $\boldsymbol{\mathcal{M}}_j^n$ refer to local Maxwellians depending only on the conservative moments. To determine $\boldsymbol{\mathcal{M}}_j^{n*}$ and $\boldsymbol{\mathcal{M}}_j^n$, we can take the conservative moments of \eqref{eq:SGS1} and \eqref{eq:SGS2}, so that the right-hand sides vanish, and the leading-order terms of the left-hand side are related only to the conservative moments. In fact, the resulting equations become the SGS method for Euler equations, which has been studied in many existing works \cite{Li2008,Hu2011}. In this sense, the scheme \eqref{eq:SGS1}\eqref{eq:SGS2} is asymptotic preserving, and therefore does not lose efficiency as $\Kn$ becomes small. However, the convergence rate of the numerical method for solving these two nonlinear equations may be affected by the choice of $\Kn$. Two numerical schemes for solving the nonlinear equations are introduced in the following subsections.

\subsection{Newton's iteration}
Since the system \eqref{eq:SGS1} or \eqref{eq:SGS2} is formed by a sequence of quadratic equations, whose Jacobian matrix can be formulated explicitly, a straightforward idea is to use Newton's method to get fast convergence. Note that both systems can be written in the following form:
\begin{equation}
      \frac{|v_{1,\bk}|}{\Delta  x} f_{\bk}-
      \frac{1}{\Kn}Q_{\bk}(\boldsymbol{f})
      =R_{\bk},
    \label{eq:Iteration_F}
\end{equation}
where the unknown vector $\boldsymbol{f}$ refers to $\boldsymbol{f}_j^{n*}$ in \eqref{eq:SGS1} and $\boldsymbol{f}_j^{n+1}$ in \eqref{eq:SGS2}, and the right-hand side $R_{\bk}$ is given by either the previous iteration or the solution in previous grid cells. For example, in \eqref{eq:SGS1}, the right-hand side $R_{\bk}$ refers to
\begin{displaymath}
    \frac{v_{1,\bk}^+}{\Delta  x}
    f_{\bk,j-1}^{n*}
    - 
    \frac{v_{1,\bk}^-}{\Delta  x}  f_{\bk,j+1}^{n}.
\end{displaymath}
Note that the schemes \eqref{eq:SGS1}\eqref{eq:SGS2} are first-order schemes. In the second-order case, the equations to be solved can still be written as \eqref{eq:Iteration_F}, but the right-hand side is slightly more complicated. Precisely speaking, the left-to-right scan will have the following definition of $R_{\bk}$:
\begin{equation}
R_{\bk}
    =-\frac{v_{1,\bk}^{-}}{\Delta x} f_{\bk,j+1}^{n}
    +\frac{v_{1,\bk}^{+}}{\Delta x} f_{\bk,j-1}^{n\ast}
    +\frac{v_{1,\bk}^{-}}{2} s_{\bk,j+1}^{n}
    -\frac{|v_{1,\bk}|}{2} s_{\bk,j}^{n}
    +\frac{v_{1,\bk}^{+}}{2} s_{\bk,j-1}^{n},
    \label{eq:L2R}
\end{equation}
and in right-to-left scan,
\begin{equation}
R_{\bk}
    =-\frac{v_{1,\bk}^{-}}{\Delta x} f_{\bk,j+1}^{n+1}
    +\frac{v_{1,\bk}^{+}}{\Delta x} f_{\bk,j-1}^{n\ast}
    +\frac{v_{1,\bk}^{-}}{2} s_{\bk,j+1}^{n\ast}
    -\frac{|v_{1,\bk}|}{2} s_{\bk,j}^{n\ast}
    +\frac{v_{1,\bk}^{+}}{2} s_{\bk,j-1}^{n\ast},
    \label{eq:R2L}
\end{equation}
Below we will focus on Newton's solver for the general system given by \eqref{eq:Iteration_F}, which can be applied to both forward and backward iterations.

The equation \eqref{eq:Iteration_F} is a system of $N^3$ equations written in the physical space. For the collision operator, the sparse representation exists only in the frequency space. Therefore, we first apply the discrete Fourier transform to write the equation \eqref{eq:Iteration_F} as  a system of the Fourier coefficients. Let 

\begin{equation} \label{eq:hatfV}
\hat{f}_{\bp} = \frac{1}{N^3} \sum_{\bk \in \{-n,\cdots,n\}^3} f_{\bk} E_{-\bp}(\bv_{\bk}), \qquad
\hat{V}_{\bp} = \frac{1}{N^3} \sum_{\bk \in \{-n,\cdots,n\}^3} |v_{1,\bk}| E_{-\bp}(\bv_{\bk}).
\end{equation}
Thus the equation \eqref{eq:Iteration_F} is equivalent to
\begin{equation}
\begin{split}
    \frac{N^3}{\Delta x}
    \sum_{\bq}
    \hat{V}_{\bp-\bq}
    \hf_{\bq}
    -
    \frac{1}{Kn}
    \sum_{\bq} \hB_{\bq,\bp-\bq}\hf_{\bq}\hf_{\bp-\bq}
   =
    \hat{R}_{\bp},
    \end{split}
    \label{eq:Iteration_F_CFT1}
\end{equation}
where $\hat{R}_{\bp}$ is the discrete Fourier transform of $R_{\bk}$ defined in the same way as \eqref{eq:hatfV}. Matrix $\hat{\bm{B}}=\left(\hB_{\bq,\bp-\bq}\right)$ is a pre calculated $N\times N$-size matrix with the element $\hB_{\bq,\bp-\bq}=\tilde{B}(\bq,\bp-\bq)-\tilde{B}(\bp-\bq,\bp-\bq)$, which can be obtained by Eq. \eqref{eq:tilde_B}. In \eqref{eq:Iteration_F_CFT1}, the range of the summation is chosen such that all the multi-indices fall in the range $\{-n,\cdots,n\}^3$.

To solve \eqref{eq:Iteration_F_CFT1}, we apply Newton's iteration which updates $\hat{f}_{\bp}$ by solving

\begin{equation}
    \sum_{\bq} a_{\bp,\bq}(\bm{\hf}^s)(\hf_{\bq}^{s+1} - \hf_{\bq}^s) =
    -\left(
    \frac{N^3}{\Delta x}
    \sum_{\bq}
    \hat{V}_{\bp-\bq}
    \hf_{\bq}^s
    -
    \frac{1}{Kn}
    \sum_{\bq} \hB_{\bq,\bp-\bq}\hf_{\bq}^s\hf_{\bp-\bq}^s
   -\hat{R}_{\bp}
    \right),
    \label{eq:FM_N1}
\end{equation}
where $s$ is the index representing the number of Newton iterations, and the coefficient matrix $\bm{A}(\bm{\hf})=\left(a_{\bp,\bq}(\bm{\hf})\right)$ is the Jacobian of the nonlinear system. Since the system is quadratic, computing the elements of $\bm{A}$ is straightforward:
\begin{equation}
    a_{\bp,\bq}(\bm{\hf}) =   
    \frac{N^3}{\Delta x}
    \hat{V}_{\bp-\bq}
    -\frac{1}{\Kn}(\hat{B}_{\bq,\bp-\bq}+\hat{B}_{\bp-\bq,\bq})\hf_{\bp-\bq}.
    \label{eq:A}
\end{equation}
By plugging \eqref{eq:A} into \eqref{eq:FM_N1}, we can reformulate Newton's iteration to the following linear system:
\begin{equation}
    \sum_{\bq} a_{\bp,\bq}(\bm{\hf}^s) \hf_{\bq}^{s+1}
    =b_{\bp}(\bm{\hf}^s) := -\frac{1}{\Kn}
    \sum_{\bq}\hat{B}_{\bp-\bq,\bq}
    \hf_{\bp-\bq}^s
    \hf_{\bq}^s
    +   \hat{R}_{\bp}.
    \label{eq:Ax_b}
\end{equation}
The detailed procedure of this method is given in Algorithm \ref{Algorithm:SGSN}, which includes the following parameters:
\begin{itemize}
\item $N_{\max}$: The maximum number of outer (SGS) iterations;
\item $S_{\max}$: The maximum number of inner (Newton) iterations;
\item $\mathit{tol}$: The threshold for the relative difference between two adjacent steps in the outer iteration;
\item $\mathit{tol}_{\mathrm{NT}}$: the threshold for the $L^2$ residual of the nonlinear equation \eqref{eq:Iteration_F_CFT1}.
\end{itemize}
The normalization at the end of each iteration is applied when the total mass needs to be specified to determine the solution (see \eqref{eq:total_mass}).

In general, the coefficient matrix $\bm{A}(\bm{\hf})$ is an $N^3 \times N^3$ dense matrix and is generally non-symmetric. Therefore, solving the linear system is highly challenging and may require a significant amount of computational time. In our implementation, we use the FGMRES solver in the PETSc package \cite{petsc-efficient,petsc-web-page,petsc-user-ref} to solve the linear system. It will be shown in Sec. \ref{sec:numerical_experiments} that the solver of the linear system takes most of the computational time in Newton's iterations. 

Interestingly, despite the theoretical quadratic convergence rate of Newton's method, the iteration still suffers from the slowdown of convergence when $\Kn$ is small. This can be seen by applying the inverse DFT to \eqref{eq:Ax_b} and rewriting the equation in the physical space, which turns out to be
\begin{equation}
\frac{|v_{1,\bk}|}{\Delta x} f_{\bk}^{s+1} - \frac{1}{\Kn} Q_{\bk}(\boldsymbol{f}^s, \boldsymbol{f}^{s+1}) - \frac{1}{\Kn} Q_{\bk}(\boldsymbol{f}^{s+1}, \boldsymbol{f}^s) = -\frac{1}{\Kn} Q_{\bk}(\boldsymbol{f}^s, \boldsymbol{f}^s) + R_{\bk}.
\label{eq:Newton_phys_space}
\end{equation}
Suppose $\Kn$ is small. The numerical solution at the previous step $\boldsymbol{f}^s$ is usually close to the local Maxwellian $\boldsymbol{\mathcal{M}}^s$:
\begin{displaymath}
\boldsymbol{f}^s = \boldsymbol{\mathcal{M}}^s + O(\Kn).
\end{displaymath}
In order to satisfy \eqref{eq:Newton_phys_space}, the new solution $\boldsymbol{f}^{s+1}$ must also satisfy
\begin{displaymath}
\boldsymbol{f}^{s+1} = \boldsymbol{\mathcal{M}}^s + O(\Kn)
\end{displaymath}
to guarantee that the $O(\Kn^{-1})$ terms in \eqref{eq:Newton_phys_space} are balanced. It appears that the method has the same behavior as the source iteration. However, since $R_{\bk}$ is unchanged during the iteration, the effective small parameter in \eqref{eq:Newton_phys_space} is $\Kn/\Delta x$. As a result, when $\Delta x$ is small, a better convergence rate is expected in practice. This will be further confirmed in our numerical experiments to be presented in Sec. \ref{sec:numerical_experiments}.

Nevertheless, solving the linear system with a full matrix could take a large amount of computational time and memory. We are therefore inspired to consider a more straightforward approach to solving the nonlinear system without solving large linear systems, which will be introduced in the next subsection.

\begin{algorithm}[!ht]
\SetAlgoLined
 
 Set the initial guess $\bm{\hf}_j^0$, $j=1,\ldots,N_x$
 
 Set $n = 0$, $\mathit{diff} = \infty$
 
 \While{$n<N_{\max}$ and $\mathit{diff} > \mathit{tol}$}
 {
    
     \For{$j=1,...,N_x$}{
       Calculate $R_{\bk}$ by \eqref{eq:L2R} \\
       Set $s = 0$ and $\bm{\hf}_j^{n,0}=\bm{\hf}_j^{n}$ \\
      \While{$s<S_{\max}$ and $\mathit{res}_{\mathrm{NT}}>\mathit{tol}_{\mathrm{NT}}$}{
        Solve the system $\bm{A}(\bm{\hf}^{n,s}) \bm{\hf}_j^{n,s+1} = \bm{b}(\bm{\hf}_j^{n,s})$ (see \eqref{eq:Ax_b}) to get $\bm{\hf}^{n,s+1}_j$

        Compute the $L^2$ residual of the nonlinear equation \eqref{eq:Iteration_F_CFT1} $\mathit{res}_{\mathrm{NT}}$ and set $s=s+1$
      }
      $\bm{\hf}_j^{n*} = \bm{\hf}_j^{n,s}$
    }
    
     \For{$j=N_x,...,1$}{
       Calculate $R_{\bk}$ by \eqref{eq:R2L} \\
       Set $s = 0$ and $\bm{\hf}_j^{n*,0}=\bm{\hf}_j^{n*}$ \\
      \While{$s<S_{\max}$ and $\mathit{res}_{\mathrm{NT}}>\mathit{tol}_{\mathrm{NT}}$}{
        Solve the system $\bm{A}(\bm{\hf}^{n*,s}) \bm{\hf}_j^{n*,s+1} = \bm{b}(\bm{\hf}_j^{n*,s})$ (see \eqref{eq:Ax_b}) to get $\bm{\hf}^{n*,s+1}_j$

        Compute the $L^2$ residual of the nonlinear equation \eqref{eq:Iteration_F_CFT1} $\mathit{res}_{\mathrm{NT}}$ and set $s=s+1$
      }
      $\bm{\hf}_j^{n+1} = \bm{\hf}_j^{n*,s}$
    }
    Compute the difference between two steps: $\mathit{diff} = \|\bm{\hf}^{n+1} - \bm{\hf}^n\|_2 / \|\bm{\hf}^n\|_2$
    
    If necessary, apply a global scaling of $\bm{\hf}_j^{n+1}$ to match the total mass in the computational domain
    
    $n=n+1$
   }
\KwResult{Fourier modes of the distribution functions $\bm{\hf}_j^n$, $j=1,\ldots,N_x$}
 \caption{Symmetric Gauss-Seidel method with Newton's iteration (SGSN)}
 \label{Algorithm:SGSN}
\end{algorithm}

\subsection{Fixed-point iteration}
Another approach to solving the nonlinear system \eqref{eq:Iteration_F} is inspired from the source iteration described in Section \ref{sec:source_iter}. The iteration reads
\begin{equation}
      \left( \frac{|v_{1,\bk}|}{\Delta  x} 
      +\epsilon_{\bk}^s\right)
      f_{\bk}^{s+1}
      =
      \epsilon_{\bk}^s
      f_{\bk}^{s}
      +
      \frac{1}{\Kn}Q_{\bk}(\boldsymbol{f}^{s})
      +R_{\bk}.
    \label{eq:Iteration_FixedPoint}
\end{equation}
The parameter $\epsilon_{\bk}^s > 0$ should be chosen to guarantee the stability of the iteration. One obvious way to select $\epsilon_{\bk}^s$ is based on the splitting of the collision term into the gain term and the loss term:
\begin{displaymath}
\mQ[f,f] = \mQ^+[f,f] - \mQ^-[f,f]
\end{displaymath}
with
\begin{equation}
    \mQ^{-}[f,f]
    =\left( \int_{\bR^3} \int_{\bS^2}
    \mB(\bv-\bv_{\ast},\sigma)f(\bv_{\ast})
    \dd \sigma\dd \bv_{\ast} \right) f(\bv),
\end{equation}
which leads to the choice
\begin{displaymath}
\epsilon_{\bk}^s = \int_{\bR^3} \int_{\bS^2}
    \mB(\bv_{\bk}-\bv_{\ast},\sigma)f^s(\bv_{\ast})
    \dd \sigma\dd \bv_{\ast}.
\end{displaymath}
In particular, for the Maxwell molecules, the kernel $\mB$ is independent of the relative velocity $\bv - \bv_{\ast}$ and
\begin{equation}
    \epsilon^{s}_{\bk}=\frac{\rho^{s}}{\Kn} \int_{\mathbb{S}^2} \mB(\sigma) \dd\sigma.
    \label{eq:epsilon}
\end{equation}

As in the method of source iteration, the fixed-point iteration also slows down as $\Kn$ decreases. However, it can be demonstrated in the linear case that the situation here is better than the source iteration. For simplicity, we assume that the collision term is linear and the parameter $\epsilon_{\bk}^s$ is independent of $\bk$ as in the case of Maxwell molecules. Furthermore, the velocity variable is assumed to be continuous, so that the scheme \eqref{eq:Iteration_FixedPoint} can be simplified to
\begin{displaymath}
  f^{s+1}(\bv) = \left( \frac{|v_1|}{\Delta x} + \frac{\kappa}{\Kn} \right)^{-1} \left( \frac{\kappa}{\Kn} + \frac{1}{\Kn} \mathcal{L} \right) f^s(\bv) + \left( \frac{|v_1|}{\Delta x} + \frac{\kappa}{\Kn} \right)^{-1} R(\bv),
\end{displaymath}
where $\epsilon$ is written as $\kappa/\Kn$ to show its relationship with the small parameter $\Kn$, and $\mathcal{L}$ denotes the linear collision operator. The convergence rate of the iteration depends on the eigenvalue of the linear operator applied to $f^s$ on the right-hand side of the scheme. Let $\lambda$ be its eigenvalue and $r(\bv)$ be the associated eigenfunction. Then
\begin{equation}
  \lambda \left( \frac{|v_1|}{\Delta x} + \frac{\kappa}{\Kn} \right) r(\bv) = \left( \frac{\kappa}{\Kn} + \frac{1}{\Kn} \mathcal{L} \right) r(\bv).
  \label{eq:eigen}
\end{equation}
When $\Kn$ is small, we apply asymptotic expansions to $\lambda$ and $r(\bv)$:
\begin{displaymath}
  \lambda = \lambda_0 + \Kn \,\lambda_1 + \cdots, \qquad r(\bv) = r_0(\bv) + \Kn \, r_1(\bv) + \cdots.
\end{displaymath}
Plugging the expansion to \eqref{eq:eigen} and balancing the $O(\Kn^{-1})$ and $O(1)$ terms on both sides, we obtain
\begin{align*}
  O(\Kn^{-1}): & \quad \lambda_0 \kappa r_0(\bv) = \kappa r_0(\bv) + \mathcal{L} r_0(\bv), \\
  O(1): & \quad \frac{|v_1|}{\Delta x} \lambda_0 r_0(\bv) + \lambda_1 \kappa r_0(\bv) + \lambda_0 \kappa r_1(\bv) = \kappa r_1(\bv) + \mathcal{L} r_1(\bv).
\end{align*}
The first equation shows that $\kappa(\lambda_0 - 1)$ is an eigenvalue of the negative semidefinite collision operator $\mathcal{L}$. By choosing $\kappa$ to be greater than the spectral radius of $\mathcal{L}$, we see that $0 < \lambda_0 \leqslant 1$, and $\lambda_0$ attains the maximum value $1$ when $\mathcal{L} r_0(\bv) = 0$, \textit{i.e.} $r_0(\bv)$ is a linearized Maxwellian:
\begin{displaymath}
  r_0(\bv) = [\boldsymbol{\phi}(\bv)]^T \boldsymbol{\alpha} \cdot \frac{1}{(2\pi)^{3/2}} \exp\left( -\frac{|\bv|^2}{2} \right),
\end{displaymath}
where $\boldsymbol{\phi}(\bv) = (1,v_1, v_2, v_3, |\bv|^2)^T$ and $\boldsymbol{\alpha} \in \mathbb{R}^5$. In this case, the $O(1)$ equation becomes
\begin{displaymath}
  \frac{|v_1|}{\Delta x} r_0(\bv) + \lambda_1 \kappa r_0(\bv) = \mathcal{L} r_1(\bv).
\end{displaymath}
To find $\lambda_1$, we take moments on both sides of the equation by multiplying $\boldsymbol{\phi}(\bv)$ and taking the integral, resulting in a system $\boldsymbol{\Sigma}(\lambda_1 \kappa) \boldsymbol{\alpha} = 0$, where
\begin{displaymath}
  \boldsymbol{\Sigma}(\lambda_1 \kappa) = \int_{\mathbb{R}^3} \left( \frac{|v_1|}{\Delta x} + \lambda_1 \kappa \right) \boldsymbol{\phi}(\bv)[\boldsymbol{\phi}(\bv)]^T \cdot \frac{1}{(2\pi)^{3/2}} \exp\left( -\frac{|\bv|^2}{2} \right) \,\mathrm{d}\bv.
\end{displaymath}
Since $\boldsymbol{\alpha} \neq 0$, we can solve the nonlinear equation $\det(\boldsymbol{\Sigma}(\lambda_1 \kappa)) = 0$ to find $\lambda_1$. The equation is quintic and has five solutions, among which the largest is
\begin{displaymath}
\lambda_{1,\max} = \frac{1}{\kappa \Delta x} \left( \sqrt{\frac{11}{24\pi}} - \sqrt{\frac{25}{8\pi}} \right) < 0.
\end{displaymath}

This indicates that the amplification factor of the fixed-point iteration is $1 - C \Kn (\kappa \Delta x)^{-1} + O(\Kn^2)$ with $C \approx 0.615$. As a comparison, the spectral radius for the source iteration is $1 - \Kn^2/2 + O(\Kn^3)$ (see \cite{Wu2020fast}). Our approach can achieve faster convergence when $\Kn$ is small.

Likewise, we summarize the procedure of the SGSFP iteration in Algorithm \ref{Algorithm:SGSFP}. 
Like SGSN, the detailed procedure of Algorithm \ref{Algorithm:SGSFP}  includes the following parameters:
\begin{itemize}
\item $N_{\max}$: The maximum number of outer (SGS) iterations;
\item $S_{\max}$: The maximum number of inner (Fixed-point) iterations;
\item $\mathit{tol}$: The threshold for the relative difference between two adjacent steps in the outer iteration;
\item $\mathit{tol}_{\mathrm{FP}}$: the threshold for the $L^2$ residual of the nonlinear equation \eqref{eq:Iteration_F}.
\end{itemize}

\begin{algorithm}[!ht]
\SetAlgoLined
 
 Set the initial guess $\bm{f}_j^0$, $j=1,\ldots,N_x$

 Set $n = 0$, $\mathit{diff} = \infty$
 
 \While{$n<N_{\max}$ and $\mathit{diff} > \mathit{tol}$}
 {
    
     \For{$j=1,...,N_x$}{
       Calculate $R_{\bk}$ by \eqref{eq:L2R} \\
       Set $s = 0$ and $\bm{f}_j^{n,0}=\bm{f}_j^{n}$ \\

      \While{$s<S_{\max}$ and $\mathit{res}_{\mathrm{FP}}>\mathit{tol}_{\mathrm{FP}}$}{
        Calculate
       $\epsilon^{n,s}_{\bk}$ by \eqref{eq:epsilon}

        Solve 
        $\left( \frac{|v_{1,\bk}|}{\Delta  x} 
      +\epsilon_{\bk}^s\right)
      f_{\bk}^{n,s+1}
      =
      \epsilon_{\bk}^{n,s}
      f_{\bk}^{n,s}
      +
      \frac{1}{\Kn}Q_{\bk}(\boldsymbol{f}^{n,s})
      +R_{\bk}$ (see \eqref{eq:Iteration_FixedPoint}) to get $\bm{f}^{n,s+1}_j$

        Compute the $L^2$ residual of the nonlinear equation \eqref{eq:Iteration_F} $\mathit{res}_{\mathrm{FP}}$ and set $s=s+1$
      }
      $\bm{f}_j^{n*} = \bm{f}_j^{n,s}$
    }
    
     \For{$j=N_x,...,1$}{
       Calculate $R_{\bk}$ by \eqref{eq:R2L} \\
       Set $s = 0$ and $\bm{f}_j^{n*,0}=\bm{f}_j^{n*}$ \\
      \While{$s<S_{\max}$ and $\mathit{res}_{\mathrm{FP}}>\mathit{tol}_{\mathrm{FP}}$}{
        Calculate
       $\epsilon^{n*,s}_{\bk}$ by \eqref{eq:epsilon}

        Solve 
        $\left( \frac{|v_{1,\bk}|}{\Delta  x} 
      +\epsilon_{\bk}^s\right)
      f_{\bk}^{n*,s+1}
      =
      \epsilon_{\bk}^s
      f_{\bk}^{n*,s}
      +
      \frac{1}{\Kn}Q_{\bk}(\boldsymbol{f}^{n*,s})
      +R_{\bk}$ (see \eqref{eq:Iteration_FixedPoint}) to get $\bm{f}^{n*,s+1}_j$

        Compute the $L^2$ residual of the nonlinear equation \eqref{eq:Iteration_F} $\mathit{res}_{\mathrm{FP}}$ and set $s=s+1$
      }
      $\bm{f}_j^{n+1} = \bm{f}_j^{n*,s}$
    }
    Compute the difference between two steps: $\mathit{diff} = \|\bm{f}^{n+1} - \bm{f}^n\|_2 / \|\bm{f}^n\|_2$
    
    If necessary, apply a global scaling of $\bm{f}_j^{n+1}$ to match the total mass in the computational domain
    
    $n=n+1$
   }
\KwResult{The distribution functions $\bm{f}_j^n$, $j=1,\ldots,N_x$}
 \caption{Symmetric Gauss-Seidel method with fixed-point iteration (SGSFP)}
 \label{Algorithm:SGSFP}
\end{algorithm}

\subsection{Boundary conditions and normalization} \label{sec:bc}

The steady-state Boltzmann equation needs appropriate boundary conditions to uniquely determine the solution. In a boundary cell, the slope in the reconstruction is defined by the one-sided difference (see \eqref{eq:slope}). In our numerical tests, two types of boundary conditions are encountered. 

For the inflow boundary conditions, the distribution function on the boundary is given for the velocity pointing into the domain. For example, the left inflow boundary condition is given by
\begin{equation} \label{eq:inflow}
f(x_{\rm L}, \bv) = g(\bv), \qquad v_1 > 0.
\end{equation}
Numerically, the boundary value $g(\bv_{\bk})$ is used to replace $f_{\bk,1/2}^L$ in \eqref{eq:reconstruction_equation} when $j = 1$. The right inflow boundary condition can be processed in a similar manner.

Another boundary condition encountered in our tests is Maxwell's wall boundary condition for solid walls. The walls are assumed to be moving at a constant velocity $\bu_W$ and have a fixed temperature $T_W$ which can affect the fluid states in the domain.  To describe the boundary condition, below we again take the left boundary as an example. Maxwell \cite{Maxwell} assumed that the gas molecules colliding with the solid wall are reflected either specularly or diffusively:
\begin{equation}
  f(x_{\rm L}, v_1, v_2, v_3) = (1-\chi) f(x_{\rm L}, -v_1, v_2, v_3) + \chi \mathcal{M}[\rho_W, \bu_W, T_W](\bv), \quad v_1 > 0,
  \label{eq:bc}
\end{equation}
where $\chi \in [0,1]$ is the accommodation coefficient specifying the proportion of diffusively reflected gas molecules, and $\mathcal{M}[\rho_W, \bu_W, T_W]$ is the local equilibrium defined by
\begin{equation} \label{eq:Maxwellian}
  \mathcal{M}[\rho_W, \bu_W, T_W](\bv) = \frac{\rho_W}{(2\pi T_W)^{3/2}} \exp \left( -\frac{|\bv - \bu_W|^2}{2T_W} \right),
\end{equation}
in which $\rho_W$ is chosen to guarantee that the mass flux is zero on the boundary:
\begin{displaymath}
\rho_W = \sqrt{\frac{2\pi}{T_W}} \int_{-\infty}^{u_{W,1}} \left( \int_{-\infty}^{+\infty}\int_{-\infty}^{+\infty} (u_{W,1}-v_1) f(v_1, v_2, v_3) \dd v_2 \dd v_3 \right) \dd v_1.
\end{displaymath}
In our test cases, we assume that the wall velocity is perpendicular to the normal direction ($u_{W,1} = 0$), so that the domain does not change due to the movement of the walls. Numerically, the values of $f_{\bk,1/2}^L$ for $v_{1,\bk} > 0$ are given by $f_{\bk,1/2}^L = 2f_{\bk}^b - f_{\bk,1/2}^R$, where
\begin{displaymath}
f_{\bk}^b = (1-\chi) f_{-\bk,1/2}^R + \chi \cdot \frac{\rho_W}{(2\pi T_W)^{3/2}} \exp \left( -\frac{|\bv_{\bk} - \bu_W|^2}{2T_W} \right),
\end{displaymath}
where $\rho_W$ is chosen such that 
\begin{displaymath}
\sum_{\bk} (v_{1,\bk} - u_{W,1}) f_{\bk}^b = 0.
\end{displaymath}

When the wall boundary conditions are imposed on both boundaries, the solution of the Boltzmann equation can be uniquely determined only if the total mass is specified by
\begin{displaymath}
\int_{x_{\rm L}}^{x_{\rm R}} \int_{\mathbb{R}^3} f(x, \bv) \,\dd \bv \,\dd x = M
\end{displaymath}
for a given $M$. Note that the symmetric Gauss-Seidel iteration does not preserve the total mass, we therefore need to apply normalization after each iteration. This is done by a uniform rescaling of all the distribution functions such that the total mass of $f^{n+1}$ equals $M$, as indicated in both algorithms.

 \section{Numerical experiments}
\label{sec:numerical_experiments}
In this section, numerical experiments are carried out to show the performance of our numerical algorithm. In all the experiments, we assume that the collision term is modeled by the Maxwell model, where $\mathcal{B}(\bv - \bv_*, \sigma)$ is a constant. According to the nondimensionalization in the appendix, the constant is chosen to be $1/(\pi\sqrt{2\pi})$. For all numerical examples, the total tolerance is $\mathit{tol}=10^{-7}$. In SGSN, Newton's method tolerance is set as $\mathit{tol}_{\mathrm{NT}}=10^{-9}$
while fixed-point's method tolerance is $\mathit{tol}_{\mathrm{FP}}=10^{-9}$ in SGSFP.

\subsection{Fourier heat flow}
Our first example is the heat conduction between parallel plates. It is assumed that both boundaries at $x_{\rm L}$ and $x_{\rm R}$ are solid walls, with temperature $T_{\rm L}$ and $T_{\rm R}$, respectively. The boundaries are fully diffusive, meaning that $\chi = 1$ in \eqref{eq:bc}. The computational domain is set to be $[0,1]$ and the total mass $M$ is set to be $1.0$. Due to the different temperatures of the plates, heat is transferred from the hot plate to the cold plate via the gas in between. In the steady state, a temperature difference can be observed between the solid wall and the gas adjacent to the wall. In general, for higher Knudsen numbers, this temperature jump is larger. Below we study the problem by considering three different choices of $T_{\rm L}$ and $T_{\rm R}$.

\subsubsection{Small temperature difference}
\label{sec:small_td}
We begin with a basic test case with $T_{\rm L} = 263.15/273.15 \approx 0.963$ and $T_{\rm R} = 283.15/273.15 \approx 1.037$. This corresponds to a $20$K temperature difference normalized about the reference temperature $273.15$K. Some numerical parameters are as follows:
\begin{itemize}
\item Spatial grid: 1D, $N_x = 30$, $\Delta x = 1/ 30$.
\item Velocity grid: 3D, $L_v = 6.62132$, $N = 17$.
\end{itemize}
Note that the discrete Fourier transforms take up only a tiny proportion of the total computational cost, and therefore it is unnecessary to choose $N$ to be the power of $2$.

\begin{figure}[!ht]
 \centering
  \subfigure[Temperature plot with $N=13$]{
	 \includegraphics[scale=.6]{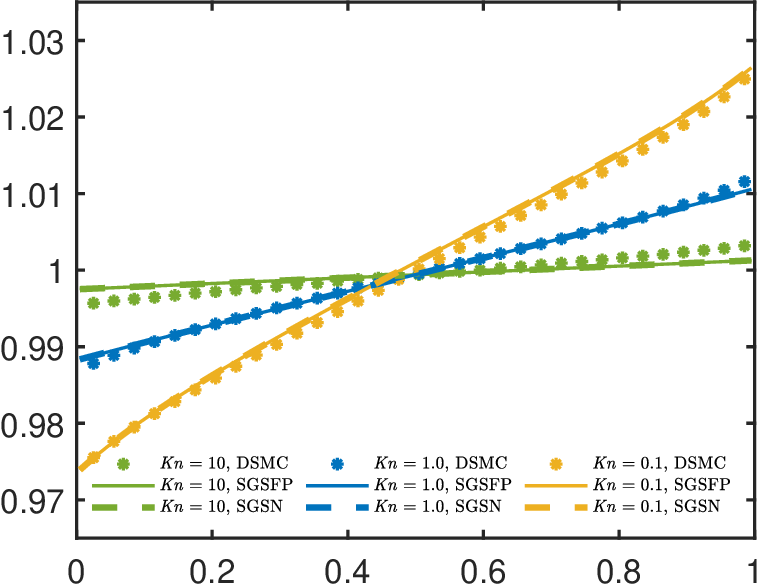}
	 \label{fig:T20_SGSFP_SGSN_N13}
	 }
 \subfigure[Temperature plot with $N=17$]{
	 \includegraphics[scale=.6]{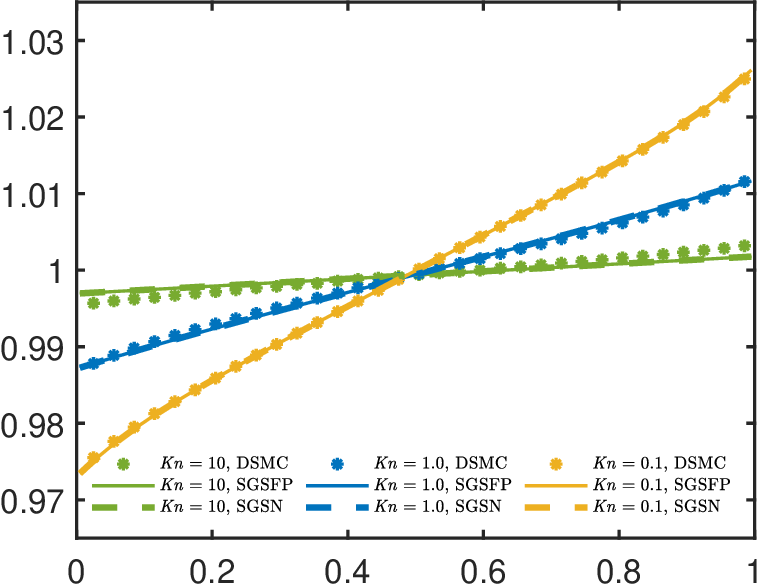}
	 \label{fig:T20_SGSFP_SGSN_N17}
	 }
    \caption{
    Temperature profiles of the Fourier flow for $T_{\rm L} = 263.15/273.15$ and $T_{\rm R} = 283.15/273.15$.}
    \label{fig:T20_SGSFP_SGSN}
\end{figure}

Our numerical results for the Knudsen numbers $0.1$, $1$ and $10$ are plotted in Figure \ref{fig:T20_SGSFP_SGSN}. We performed the same simulation using the DSMC method, and the results are also provided as a reference. Due to the temperature jump, the temperature difference between both sides is less than $T_{\rm R} - T_{\rm L}$ ($\approx 0.073$), and the temperature profile approaches a constant when $\Kn$ gets larger. Results of SGSFP and SGSN agree well with each other for all Knudsen numbers, since the only difference between these methods is the nonlinear solver of  \eqref{eq:Iteration_F}.

Compared with DSMC results, our method produces similar curves for $\Kn = 0.1$ and $\Kn = 1.0$. However, a larger discrepancy can be observed for $\Kn = 10$. In general, when the gases are more rarefied, the distribution function becomes more irregular. In particular, discontinuities may appear in the distribution functions for points close to the boundary of the domain. As a result, it is harder for the spectral method to capture them accurately, resulting in worse approximations of the overall solution. The approximation can be improved by increasing $N$, which will be demonstrated soon.

The two methods SGSFP and SGSN can validate each other, but the computational cost is quite different. Table \ref{table:time_T20} records the running time of both methods. It is seen that the SGSFP method is significantly faster, and the reason for the slowness of SGSN is the high computational cost of the linear solver, which takes up more than 99\% of the computational time. The number of iterations for both methods is generally the same, since the difference between the two methods only affects inner iterations. The number of iterations increases slightly for smaller Knudsen numbers. For the SGSFP method, the average computational time $T_{\mathrm{avg}}$ increases for decreasing $\Kn$, due to the increasing number of inner iterations. For the SGSN method, although $T_{\mathrm{avg}}$ also rises for denser gases, it generally shows a more stable behavior due to the superiority of Newton's iteration compared with the fixed-point iteration. We believe Newton's method may achieve better performance if accompanied by a more carefully designed linear solver, but at the moment, we will mainly focus on the SGSFP method below.

\begin{table}[!ht]
\caption{The computational time for the Fourier flow for $T_{\rm L} = 263.15/273.15$ and $T_{\rm R} = 283.15/273.15$. $T_{\mathrm{total}}$: total computational time; $T_{\mathrm{lin}}$: computational time spent on the linear solver; $N_{\mathrm{iter}}$: number of iterations; $T_{\mathrm{avg}}$: average computational time for each iteration.}
\centering
\begin{tabular}{c@{\qquad\qquad}c@{\qquad\qquad}ccc@{\qquad\qquad}cccc}
\hline
& & \multicolumn{3}{c@{\qquad\qquad}}{SGSFP} & \multicolumn{4}{c}{SGSN} \\ \hline
& & $T_{\mathrm{total}}$ & $N_{\mathrm{iter}}$ & $T_{\mathrm{avg}}$
& $T_{\mathrm{total}}$ & $T_{\mathrm{lin}}$ & $N_{\mathrm{iter}}$ & $T_{\mathrm{avg}}$ \\
\hline
& $\Kn=0.1$         & 523s & 19 &  28s          &37333s&36989s   &19 & 1965s\\ 
$N = 13$ & $\Kn=1$           & 88s & 8 &  11s     
&14466s & 14334s &8 & 1808s \\ 
& $\Kn=10$         & 45s & 6 & 8s &10415s & 10313s& 6     & 1736s       \\ \hline  
& $\Kn=0.1$         & 2396s & 18 &  133s           &395499s & 393636s  &18 &21972s             \\  
$N=17$ & $\Kn=1$           & 407s & 8 &  51s      
&160096s& 159346s &8 & 20012s\\ 
& $\Kn=10$         & 187s & 6 &  31s &113444s       & 112897s&6      &  18907s         \\ \hline  
\end{tabular}
\label{table:time_T20}
\end{table}

\begin{figure}[!htbp]
 \centering
 \subfigure[Small Knudsen numbers]{
	 \includegraphics[width=0.44\linewidth]{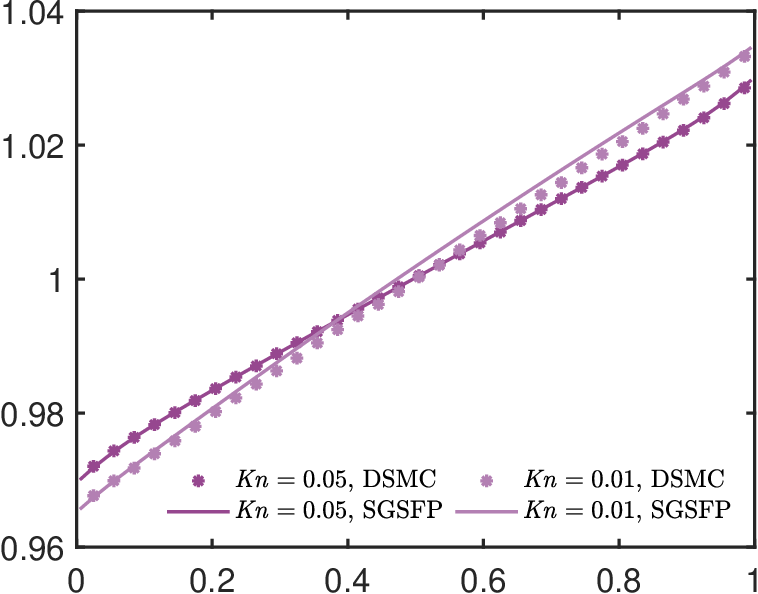}
	 \label{fig:T20_Kn}
	 }
   \subfigure[Large Knudsen numbers]{
	 \includegraphics[width=0.44\linewidth]{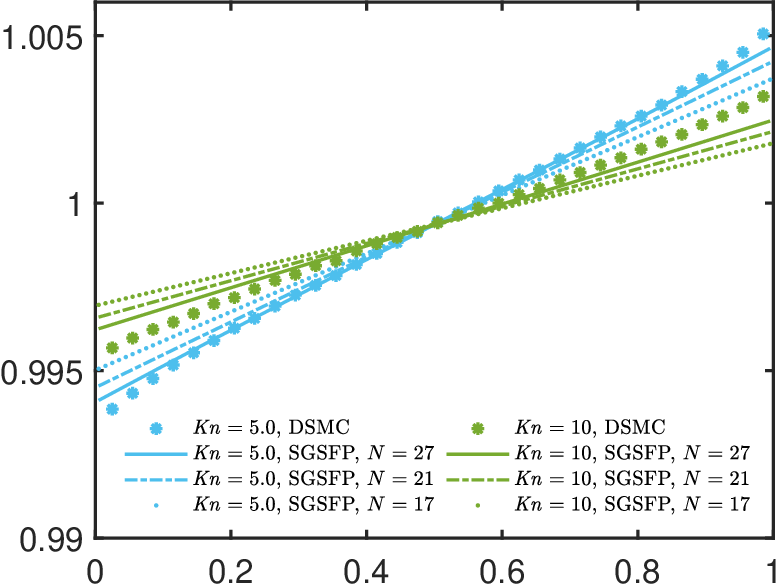}
	 \label{fig:T20_N}
	 }
    \caption{
    Temperature profiles of the Fourier flow for $T_{\rm L}=263.15/273.15$ and $T_{\rm R}=283.15/273.15$ at small Knudsen numbers (left) and large Knudsen numbers (right).}
	\label{fig:T20}
\end{figure}

More results with smaller Knudsen numbers or larger values of $N$ are given in Figure \ref{fig:T20}, all of which are simulated using the SGSFP method. Note that for Knudsen numbers $5.0$ and $10.0$, we have increased the number of Fourier modes to achieve better results. For small $\Kn$, the major difficulty is the convergence of the nonlinear algebraic problem, whereas for large $\Kn$, the major difficulty comes from the high computational complexity of the collision term ($O(N^6)$ in the current simulation) and the need for a larger number of Fourier modes. When $\Kn = 10$, the result is unsatisfactory even for $N = 27$, which manifests the drawback of the spectral method under this circumstance.

\begin{figure}[!htbp]
 \centering
 \subfigure[$g_1(x,v_1)$]{
	 \includegraphics[width=0.44\linewidth]{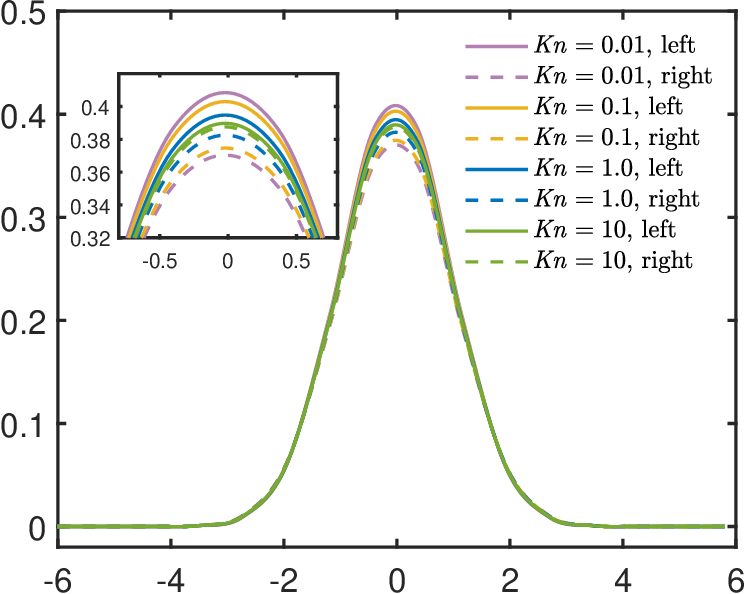}
	 \label{fig:T20_g1}
	 }
   \subfigure[$g_2(x,v_2)$]{
	 \includegraphics[width=0.44\linewidth]{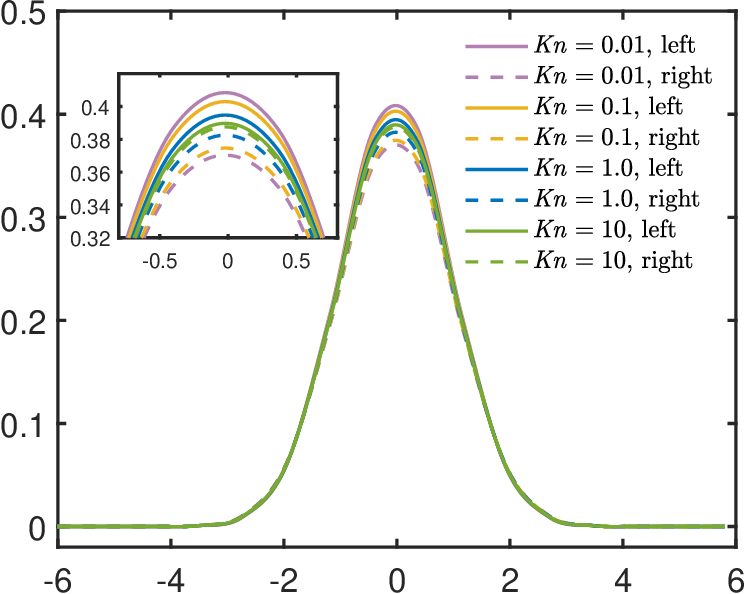}
	 \label{fig:T20_g2}
	 }
    \caption{The one-dimensional marginal distribution function $g_1(x,v_1)$
    and $g_2(x,v_2)$ of Fourier heat flow for $T_{\rm L}=263.15/273.15$ and $T_{\rm R}=283.15/273.15$.
    }
	\label{fig:marginal_T20_1D}
\end{figure}

In Figure \ref{fig:marginal_T20_1D}, we plot the marginal distribution functions on both boundaries of the domain, defined by
\begin{displaymath}
     g_1(x,v_1)=\int_{\bR^2} f(x,v_1,v_2,v_3) \,\dd v_2  \,\dd v_3,
     \qquad
    g_2(x,v_2)=\int_{\bR^2} f(x,v_1,v_2,v_3) \,\dd v_1 \,\dd v_3.
\end{displaymath}
It can be observed that for larger Knudsen numbers, the distribution functions on both sides are closer to each other. In fact, in the collisionless case, the distribution function $f(x,\bv)$ should be independent of $x$ \cite{Cai2019Convergence}.
For small Knudsen numbers, the Boltzmann equation behaves more like the Euler equations, in which the pressure is a constant, and therefore the distribution function on the high-temperature side has small values around velocity zero, indicating a smaller density to balance the pressure.

The computational time for $N = 17$ is summarized in Tab. \ref{table:time_T20_smallKn}. As $\Kn$ decreases, the numbers of both outer (SGS) and inner (fixed-point) iterations increase. Note that different Knudsen numbers lead to different solutions, so the same stopping criterion may actually mean different absolute accuracy in different tests. It is therefore inappropriate to make a direct comparison for the number of outer iterations $N_{\mathrm{iter}}$. Nevertheless, it is encouraging to notice that from $\Kn = 10$ to $\Kn = 0.01$, the average computational time for each iteration increases only by about 20 times, indicating that our method does not worsen too fast when $\Kn$ decreases.

\subsubsection{Large temperature difference}

\begin{figure}[!htbp]
 \centering
 \subfigure[$T_{\rm L}=223.15/273.15$,\quad $T_{\rm R}=323.15/273.15$]{
	 \includegraphics[width=0.44\linewidth]{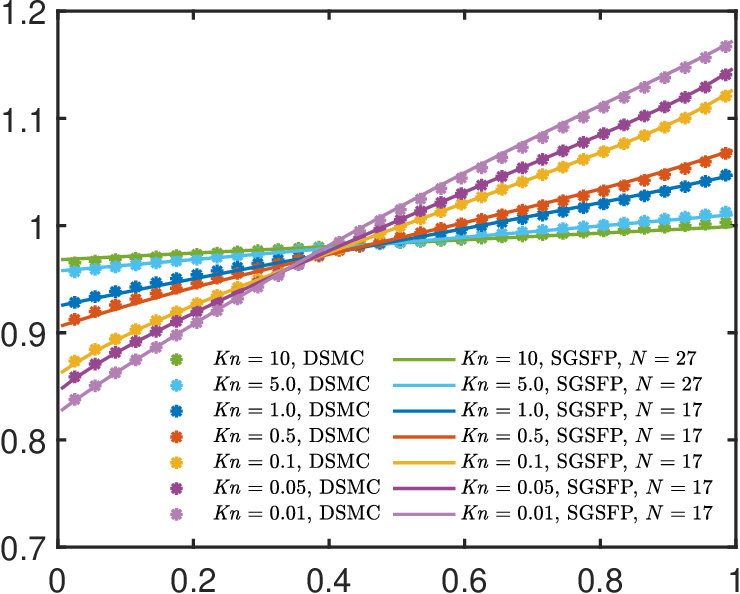}
	 \label{fig:T100}
	 }
   \subfigure[$T_{\rm L}=1$,\quad $T_{\rm R}=4$]{
	 \includegraphics[width=0.44\linewidth]{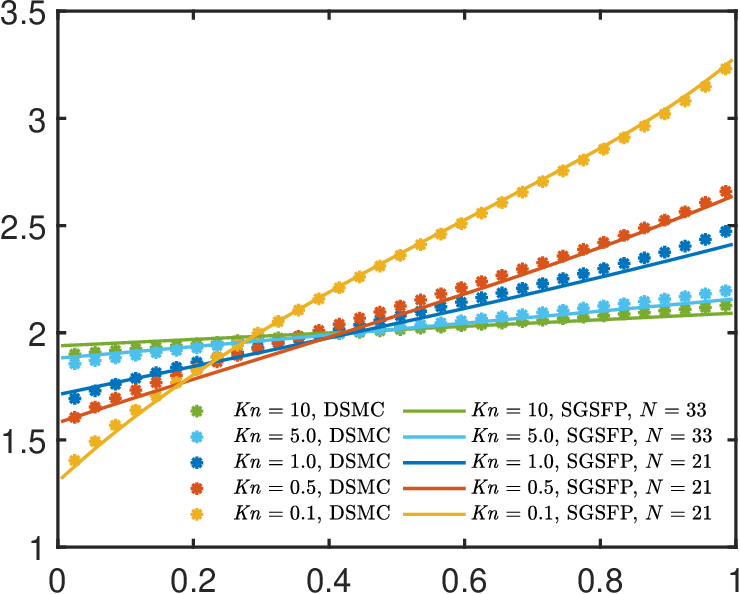}
	 \label{fig:F4}
	 }
 	\caption{
   The temperature profiles for Fourier flows with large temperature differences. }
	\label{fig:T100_F4}
\end{figure}

We now increase the temperature difference to $100$K, so that $T_{\rm L} = 223.15/273.15$ and $T_{\rm R} = 323.15/273.15$. The velocity domain is again chosen as $[-L_v, L_v]^3$ with $L_v = 6.62132$. The temperature profiles for different Knudsen numbers are plotted in Fig. \ref{fig:T100}, and the DSMC results are again given as reference solutions. A good agreement between our solution and the DSMC solution can be observed up to $\Kn = 1.0$ with $N = 17$. However, for Knudsen numbers $5$ and $10$, the difference between two sets of solutions is still significant even with $N = 27$, since a larger temperature difference between walls leads to a stronger discontinuity in the distribution function. The computational cost and the number of iterations are given in Tab. \ref{table:time_T100_smallKn}, which shows the same trend as in Sec. \ref{sec:small_td}.

A more challenging test case with a temperature ratio $1:4$ is also done in our experiments (see Fig. \ref{fig:F4}). Due to the large temperature of the right plate, we increase the velocity domain to $[-2L_v, 2L_v]^3$ in the simulation. Meanwhile, a larger value of $N$ is adopted for the velocity discretization. All the results are qualitatively correct, but quantitatively accurate results require more Fourier modes except for $\Kn = 0.1$. Note that for such a large temperature ratio, we did not carry out simulations for smaller Knudsen numbers $\Kn = 0.01$ and $\Kn = 0.05$, since the inner iteration takes too much time due to both the small Knudsen numbers and the large value of $N$.

\begin{table}[!ht]
\caption{The computational time for the Fourier flow with $T_{\rm L}=263.15/273.15$ and $T_{\rm R}=283.15/273.15$. All results are for $N = 17$. $T_{\mathrm{total}}$: total computational time; $N_{\mathrm{iter}}$: number of iterations; $T_{\mathrm{avg}}$: average computational time for each iteration. }
\centering
\begin{tabular}{ccccccc}
\hline
& $\Kn = 0.01$
& $\Kn = 0.05$
& $\Kn = 0.1$
& $\Kn = 1$
& $\Kn = 5$
& $\Kn = 10$ \\
\hline
$T_{\mathrm{total}}$ & 71740s & 5440s & 2396s & 407s &218s& 187s \\
$N_{\mathrm{iter}}$ & 116 & 28 & 18 & 8 &6& 6 \\
$T_{\mathrm{avg}}$ & 618s & 194s & 133s & 51s &36s& 31s \\
\hline
\end{tabular}
\label{table:time_T20_smallKn}
\end{table}

\subsubsection{Computational cost}
\begin{figure}[!htbp]
 \centering
   
 \includegraphics[width=0.44\linewidth]{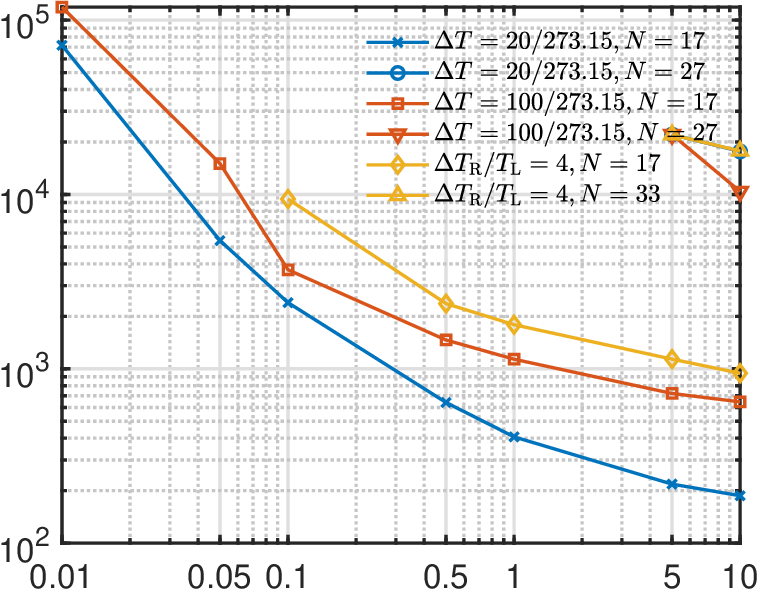}
 
    \caption{Computational time of the Fourier flow.    }
	\label{pic:computational_cost_Fourier_flow}
\end{figure}

The computational time for the simulations using the SGSFP method is summarized in Fig. \ref{pic:computational_cost_Fourier_flow}. In these tests, no parallelization is applied. As expected, the computational time increases for smaller $\Kn$ or larger $N$. To improve the computational efficiency for large Knudsen numbers, we need to improve the time complexity for the calculation of collision operators. This includes the application of fast spectral methods \cite{Hu2017} or adaptive methods \cite{CaiBurnett2022Adaptive}. To reduce the computational cost for small Knudsen numbers, one can turn to the method of GSIS \cite{Wu2020fast} or directly apply moment models such as the regularized 13-moment equations \cite{struchtrup2003regularization}.

\begin{table}[!ht]
\caption{The computational time for the Fourier flow with $T_{\rm L}=223.15/273.15$ and $T_{\rm R}=323.15/273.15$. All results are for $N = 17$. $T_{\mathrm{total}}$: total computational time; $N_{\mathrm{iter}}$: number of iterations; $T_{\mathrm{avg}}$: average computational time for each iteration.}
\centering
\begin{tabular}{cccccccc}
\hline
& $\Kn = 0.01$
& $\Kn = 0.05$
& $\Kn = 0.1$
& $\Kn = 0.5$
& $\Kn = 1$
& $\Kn = 5$
& $\Kn = 10$ \\
\hline
$T_{\mathrm{total}}$ &118669s &15049s &3698s &1463s  &1136s &722s&646s  \\
$N_{\mathrm{iter}}$ & 143 & 32 &20  &11&10  &8& 8 \\
$T_{\mathrm{avg}}$ &830s  &470s  &185s  &133s  &114s&90s&81s  \\
\hline
\end{tabular}

\label{table:time_T100_smallKn}
\end{table}

\subsection{Couette flow}
The Couette flow, which is another benchmark channel flow driven by the movement of the walls, is simulated in our tests. The gas is between two parallel plates with the same temperature, but the plates move in opposite directions at the same velocity $u_{W,2}$, leading to a flow velocity parallel to the plates in the steady-state solution. For rarefied gases, the velocity of the fluid is always smaller than the velocity of the wall, which is known as the velocity slip in shear flows.

\begin{figure}[!htbp]
 \centering
  \subfigure[$\Kn=0.01$, $N=17$]{
	 \includegraphics[width=0.3\linewidth]{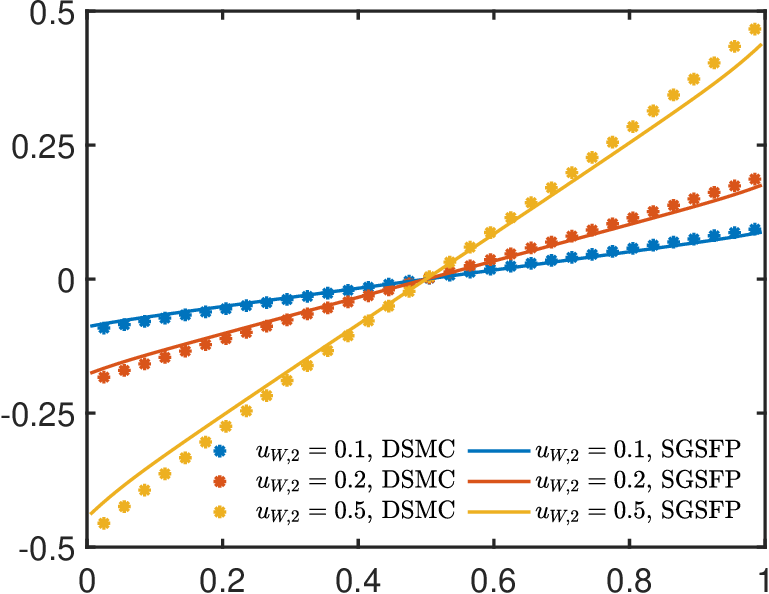}
	 \label{fig:C_Kndot01}
	 }
   \subfigure[$\Kn=0.05$, $N=17$]{
	 \includegraphics[width=0.3\linewidth]{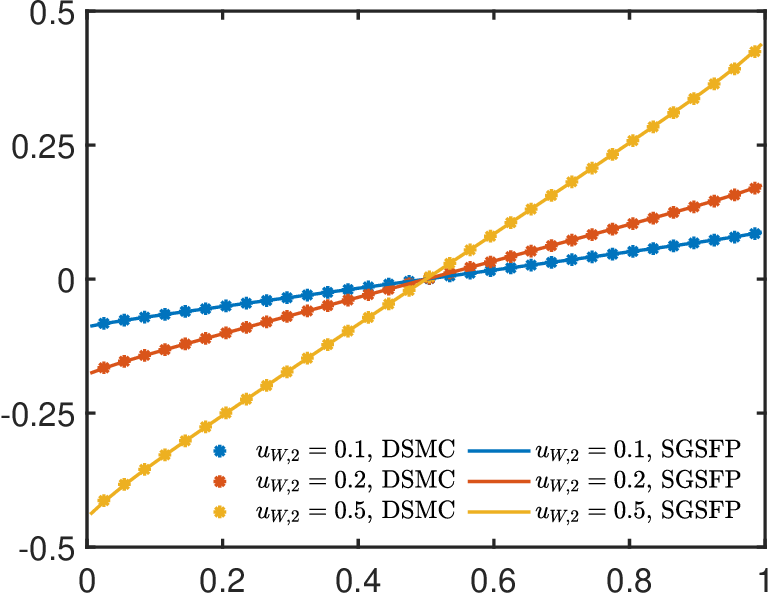}
	 \label{fig:C_Kndot05}
	 }
 \subfigure[$\Kn=0.1$, $N=17$]{
	 \includegraphics[width=0.3\linewidth]{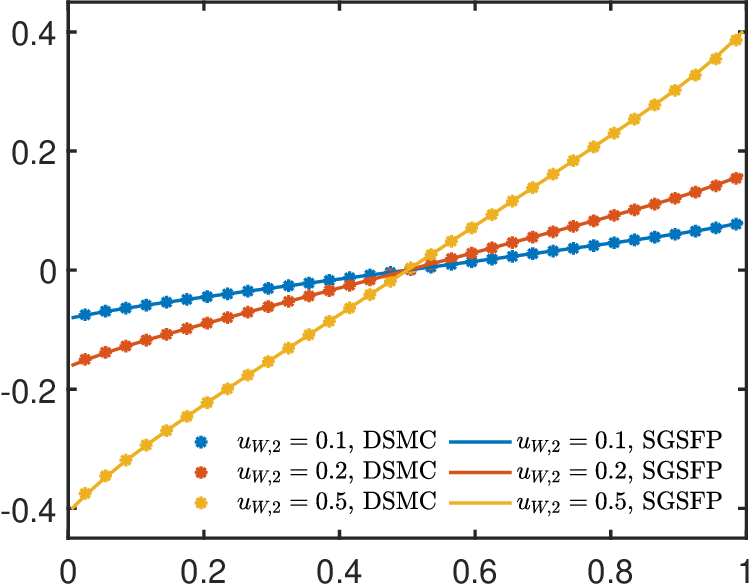}
	 \label{fig:C_Kndot1}
	 }
   \subfigure[$\Kn=0.5$, $N=17$]{
	 \includegraphics[width=0.3\linewidth]{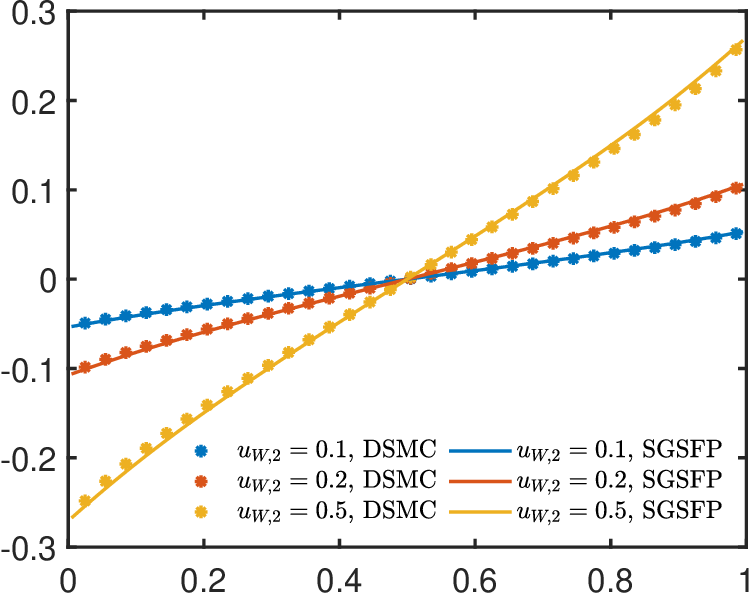}
	 \label{fig:C_Kndot5}
	 }
   \subfigure[$\Kn=1.0$, $N=17$]{
	 \includegraphics[width=0.3\linewidth]{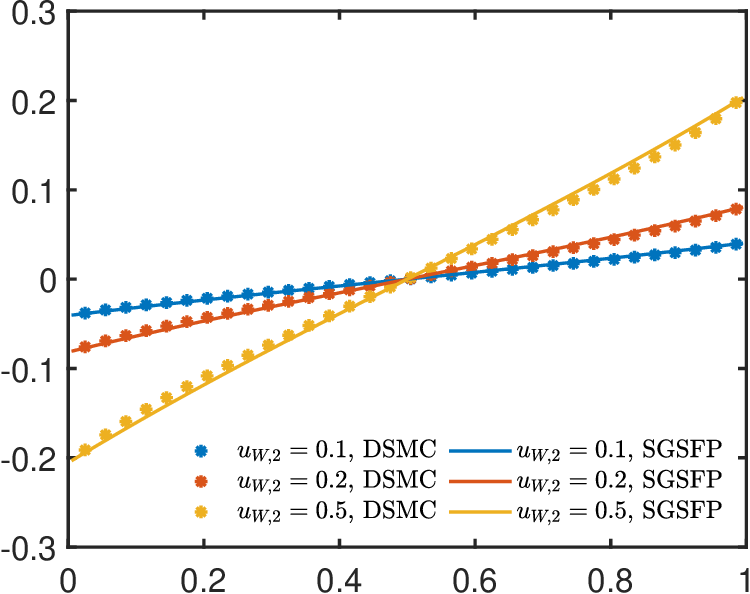}
	 \label{fig:C_Kn1}
	 }
   \subfigure[$\Kn=10$, $N=33$]{
	 \includegraphics[width=0.3\linewidth]{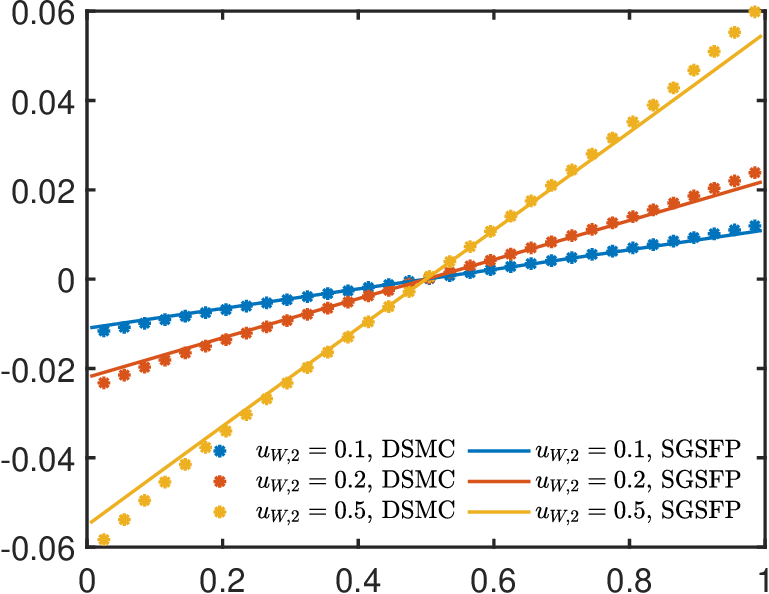}
	 \label{fig:C_Kn10}
	 }
    \caption{The velocity curves of Couette flows with different Knudsen numbers and wall speeds.}
    \label{fig:Couette}
\end{figure}

In our experiments, we again set the spatial domain to be $[0,1]$, and the velocity space to be $[-L_v, L_v]^3$ with $L_v = 6.62132$. Three different wall speeds $u_{W,2} = 0.1$, $0.2$ and $0.5$ are studied. 
The numerical results of the SGSFP method for six Knudsen numbers, as well as the reference solution given by the DSMC method, are shown in Fig. \ref{fig:Couette}, and the agreement of two sets of solutions again validates our algorithm. For $\Kn = 0.01$, the speed of the fluid on both boundaries is nearly the same as the speed of the wall, whereas in the rarefied case $\Kn = 10$, a large velocity slip is observed.

\begin{figure}[!htbp]
 \centering
 \subfigure[ $u_{W,2}=0.1$ ]{
	 \includegraphics[width=0.3\linewidth]{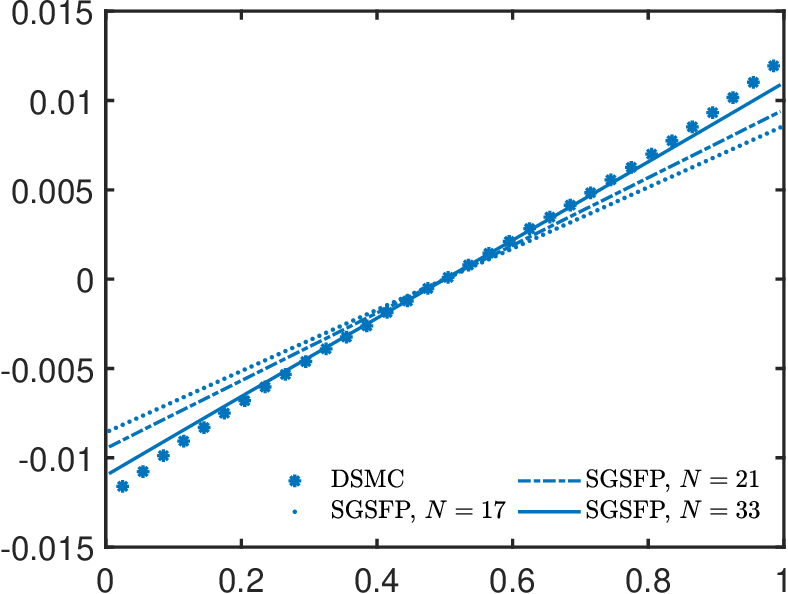}
	 }
   \subfigure[$u_{W,2}=0.2$  ]{
	 \includegraphics[width=0.3\linewidth]{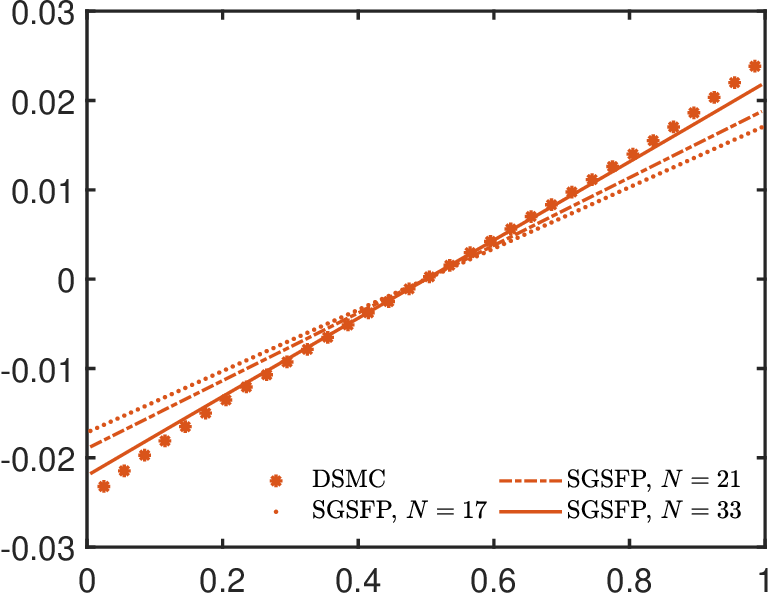}
	 }
   \subfigure[ $u_{W,2}=0.5$ ]{
	 \includegraphics[width=0.3\linewidth]{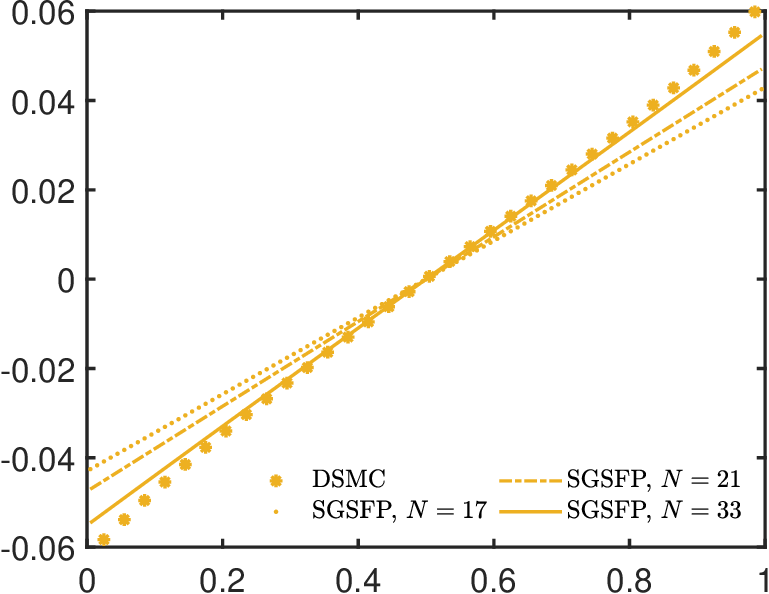}
	 }
 \caption{The velocity curves of Couette flows for different wall speeds at $\Kn = 10$.}
	\label{fig:Couette_Kn10_N}
\end{figure}

As in the case of Fourier flows, for $\Kn = 10$, a larger $N$ is used in the numerical simulations, compared with all other Knudsen numbers ranging from $0.01$ to $1.0$, which is again due to the decreasing regularity of the distribution functions as $\Kn$ increases. The convergence with respect to $N$ is demonstrated in Fig. \ref{fig:Couette_Kn10_N}. By comparing the results of $N = 17$ and $N = 33$, we see that the spectral method still shows fast convergence despite the large Knudsen number.

\begin{figure}[!htbp]
 \centering
 \subfigure[ $u_{W,2}=0.1$, left boundary]{
	 \includegraphics[width=0.3\linewidth]{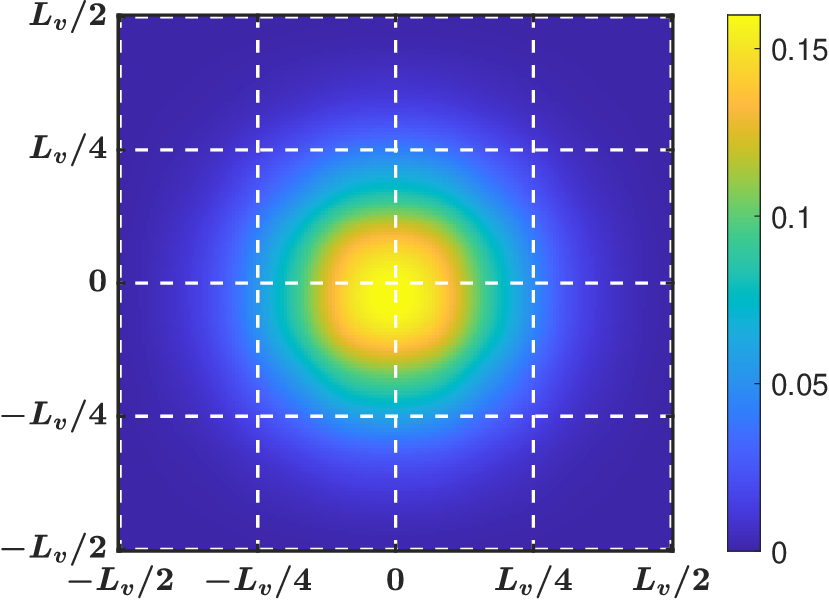}
	 }
   \subfigure[$u_{W,2}=0.2$, left boundary]{
	 \includegraphics[width=0.3\linewidth]{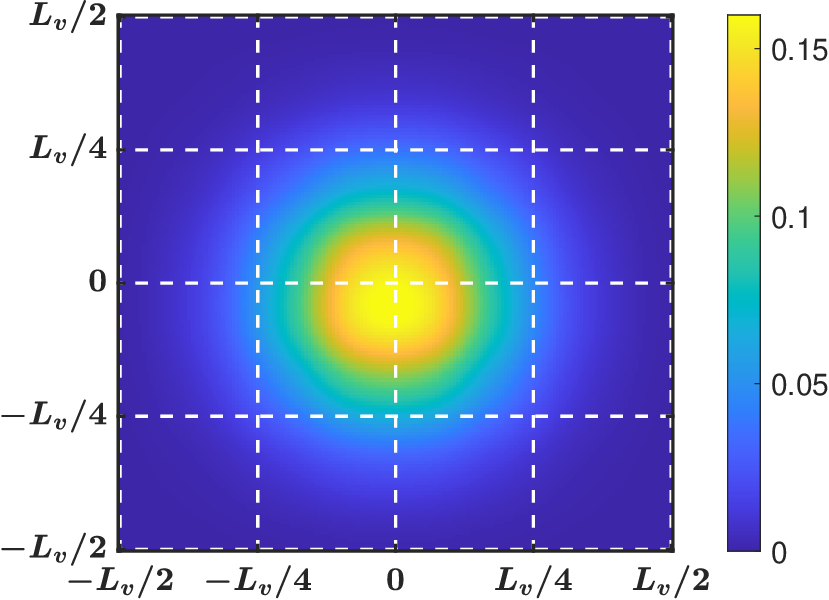}
	 }
   \subfigure[ $u_{W,2}=0.5$, left boundary]{
	 \includegraphics[width=0.3\linewidth]{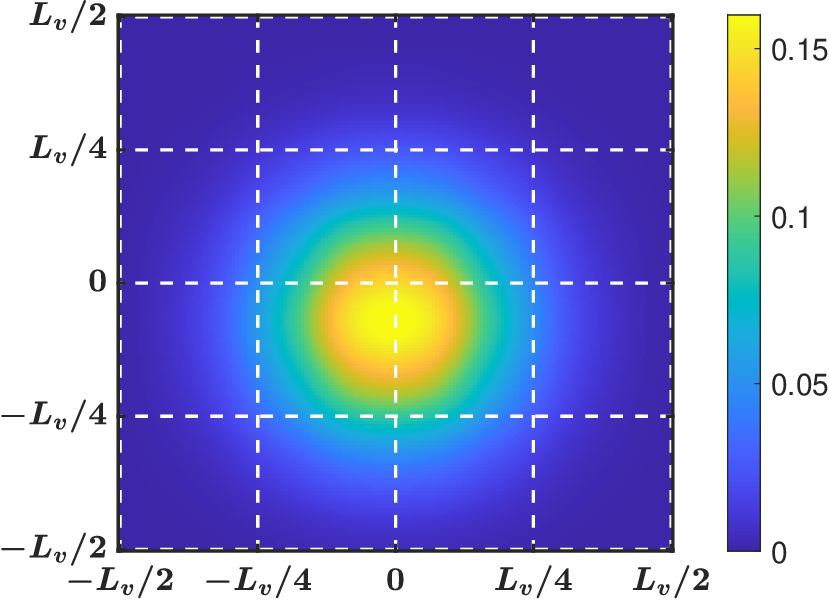}
	 }\\
 \subfigure[ $u_{W,2}=0.1$, right boundary]{
	 \includegraphics[width=0.3\linewidth]{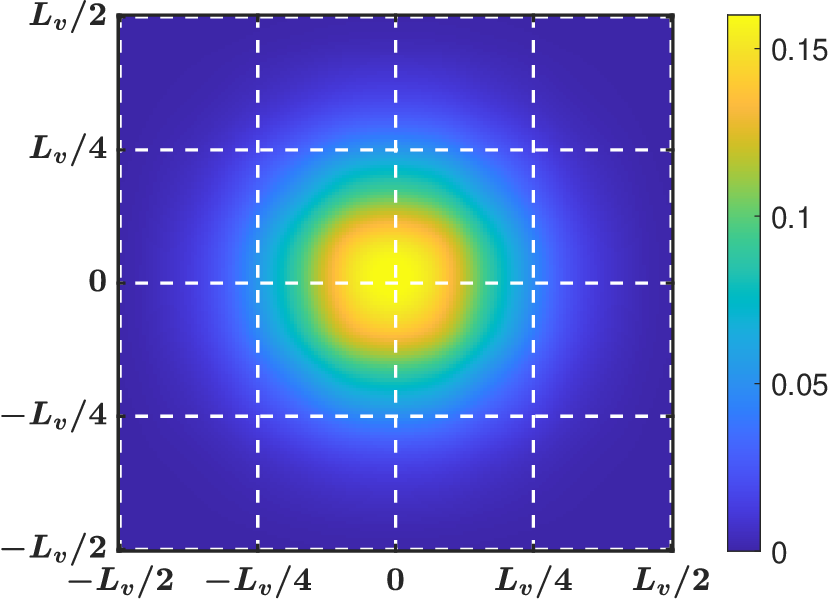}
	 }
   \subfigure[$u_{W,2}=0.2$, right boundary]{
	 \includegraphics[width=0.3\linewidth]{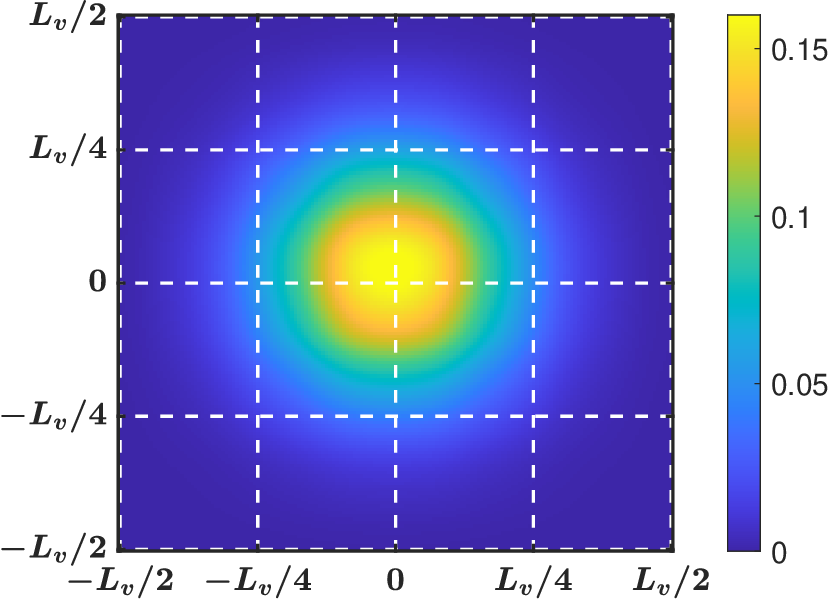}
	 }
   \subfigure[ $u_{W,2}=0.5$, right boundary]{
	 \includegraphics[width=0.3\linewidth]{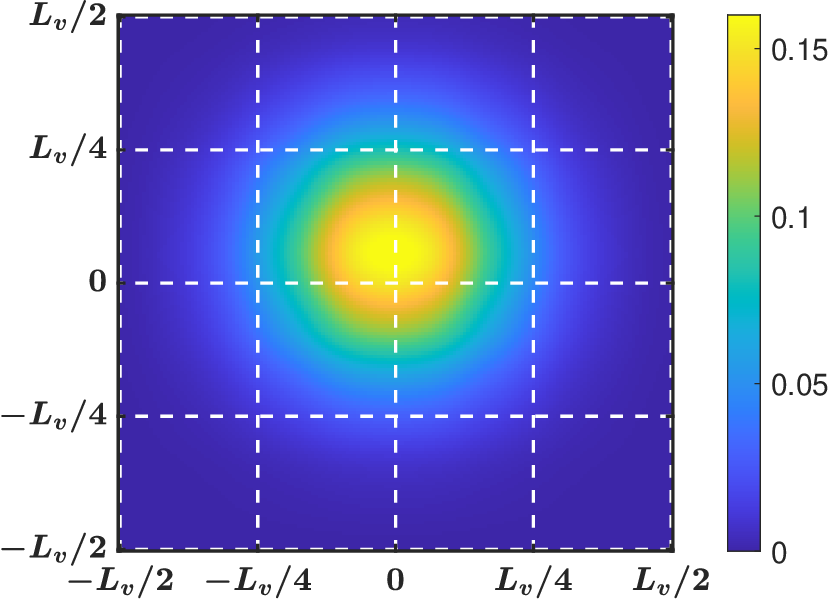}
	 }
 \caption{The two-dimensional marginal distribution function $g(x,v_1,v_2)$ of Couette flow with $\Kn = 0.1$ on boundaries of the domain.}
	\label{fig:Couette_marginal}
\end{figure}

\begin{figure}[!htbp]
 \centering
  
	 \includegraphics[width=0.44\linewidth]{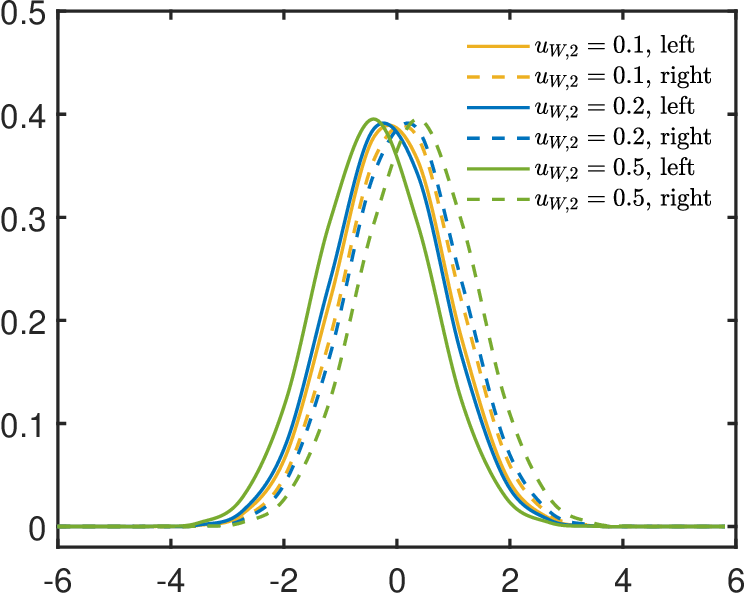}
	 \label{fig:Couette_g2}
	
    \caption{The one-dimensional marginal distribution function $g_2(x,v_2)$ of Couette flow at $\Kn=0.1$.
    }
	\label{fig:marginal_C_1D}
\end{figure}

Fig. \ref{fig:Couette_marginal} shows the marginal distribution functions $g(x,v_1,v_2)$ defined by
\begin{displaymath}
  g(x,v_1,v_2) = \int_{\mathbb{R}} f(x,v_1,v_2,v_3) \,\mathrm{d}v_3
\end{displaymath}
on both boundaries of the domain. All the figures in the top row have focal points slightly below the origin, indicating the negative velocity on the left boundary, whereas the figures in the bottom row are the opposite. The figures correctly show the symmetry of the flow, and it can be noticed that the distribution functions are further away from the center for larger $u_{W,2}$. This can be observed more clearly in Fig. \ref{fig:marginal_C_1D}, where one-dimensional marginal distribution functions are plotted.

The computational time for these simulations is given in Tab. \ref{tab:computational_cost_Couette_flow}. For the fixed $\Kn$ and $N$, the computational cost is nearly the same for different wall velocities. The change of computational time with $\Kn$ again exhibits the same behavior as in the tests of Fourier flows.

\begin{table}[ht]
\caption{Computational time of the Couette flow. The last column is separated due to a different value of $N$.}
\begin{tabular}{c||ccccc|c}
\hline
Knudsen number & $\Kn=0.01$  & $\Kn=0.05$  & $\Kn=0.1$ &$\Kn=0.5$  & $\Kn=1$    & $\Kn=10$ \\
Velocity discretization           &  $N=17$       &    $N=17$      &    $N=17$   & $N=17$   & $N=17$    & $N=33$              \\ 
\hline
$u_{W,2}=0.1$ & 117948s&9287s&4486s&1034s&935s&32042s\\
$u_{W,2}=0.2$ & 136909s&10474s&4628s&1129s&1000s&11754s\\
$u_{W,2}=0.5$ & 128234s&11348s&4811s&1197s&1028s&14158s\\
\hline
\end{tabular}
\label{tab:computational_cost_Couette_flow}
\end{table}

\subsection{Shock structures}
The computation of shock structures is another benchmark test for rarefied gas flows, which requires solving the steady-state solution on the unbounded domain $\mathbb{R}$. According to the Rankine-Hugoniot condition, for a stationary shock wave with Mach number $\Ma$, the fluid state before the shock is given by an equilibrium with the following macroscopic variables:
\begin{equation} \label{eq:left}
    \rho_{\rm L}=1,\qquad \bu_{\rm L}=\left( \sqrt{\frac{5}{3}}{\Ma}, 0, 0 \right)^T,\qquad
    T_{\rm L}=1,
\end{equation}
and the fluid after the shock is in an equilibrium with
\begin{equation} \label{eq:right}
    \rho_{\rm R} = \rho_{\rm L}\frac{4{\Ma}^2}{{\Ma}^2+3},\quad
    \bu_{\rm R} = \left(\sqrt{\frac{5}{3}}\frac{{\Ma}^2+3}{4{\Ma}}, 0, 0\right)^T,\quad
    T_{\rm R} = \frac{(5{\Ma}^2-1)( {\Ma} ^2+3)}{16{\Ma}^2}.
\end{equation}
Since the domain is unbounded, the Knudsen number is only a scaling of the spatial variable, which does not change the nature of the solution. Here we simply set $\Kn = 1$. Numerically, we set the computational domain to be $[-20,20]$, and we apply the inflow boundary condition on the left boundary as in \eqref{eq:inflow}, where $g(\bv)$ is chosen to be the local equilibrium $\mathcal{M}[\rho_{\rm L}, \bu_{\rm L}, T_{\rm L}](\bv)$ (see \eqref{eq:Maxwellian} for the definition). The right boundary condition is imposed symmetrically. Again, we only use SGSFP in the simulations below.

\begin{figure}[!htbp]
 	\centering
 	\subfigure[Normalized density $\bar{\rho}$, velocity $\bar{u}$ and temperature $\bar{T}$]{
 		\includegraphics[width=0.45\linewidth, height=0.3\linewidth]{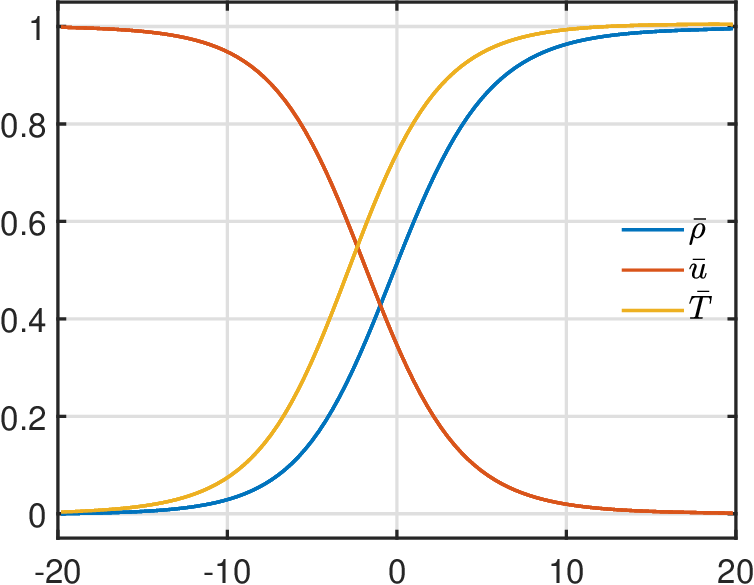} 
         \label{fig:Ma1dot55}}
         \subfigure[stress $\sigma$ and heat flux $q$]{
 		\includegraphics[width=0.45\linewidth, height=0.3\linewidth]{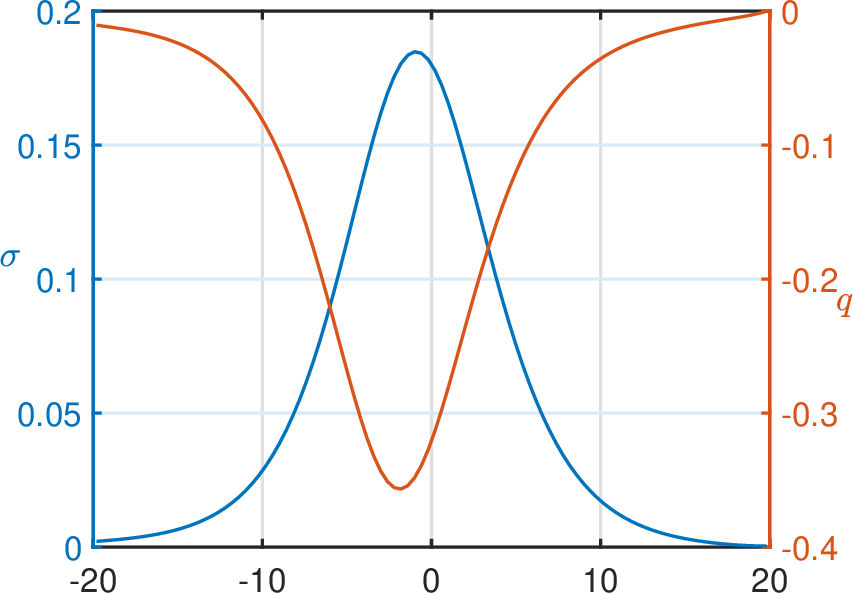} 
         \label{fig:Ma1dot55_sigma_q}}
 	\caption{Shock structure for Mach number $\Ma=1.55$.}
	\label{fig:Shock_Ma1.55}
\end{figure}

\begin{figure}[!htbp]
      \centering
 	\subfigure[Normalized density $\bar{\rho}$, velocity $\bar{u}$ and temperature $\bar{T}$]{
	 \includegraphics[width=0.45\linewidth, height=0.3\linewidth]{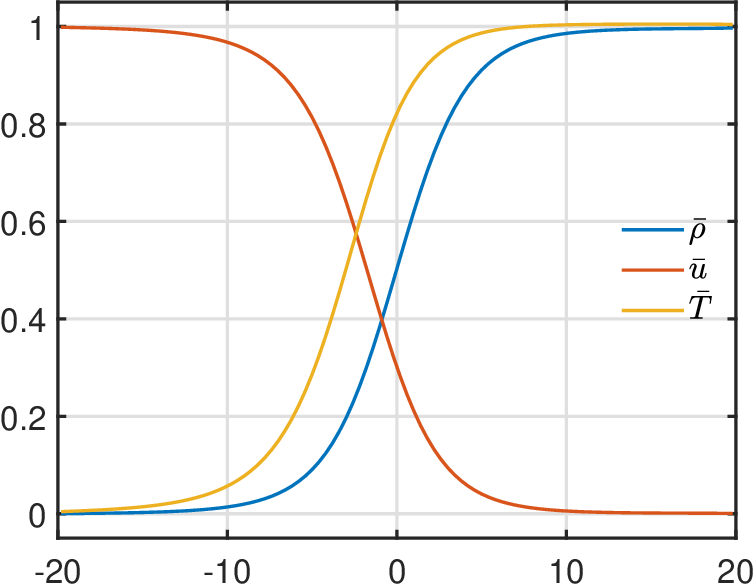}
        \label{fig:Ma2}
	 }	
       \subfigure[stress $\sigma$ and heat flux $q$]{
\includegraphics[width=0.45\linewidth, height=0.3\linewidth]{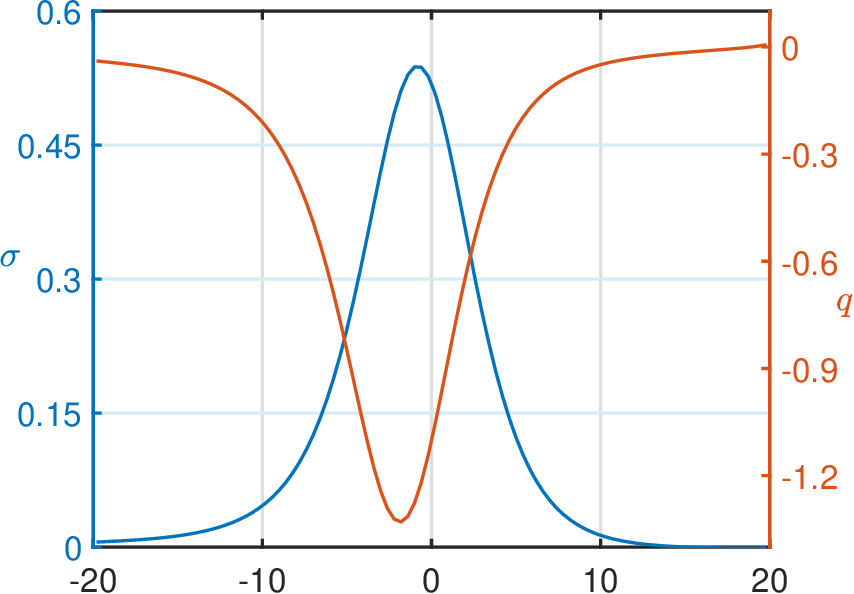}
        \label{fig:Ma2_sigma_q}
	 }
  \caption{Shock structures with Mach number $\Ma=2$.}
	\label{fig:Shock_Ma2}
\end{figure}

\begin{figure}[!htbp]
    \centering
 	\subfigure[Normalized density $\bar{\rho}$, velocity $\bar{u}$ and temperature $\bar{T}$]{
	 \includegraphics[width=0.45\linewidth, height=0.3\linewidth]{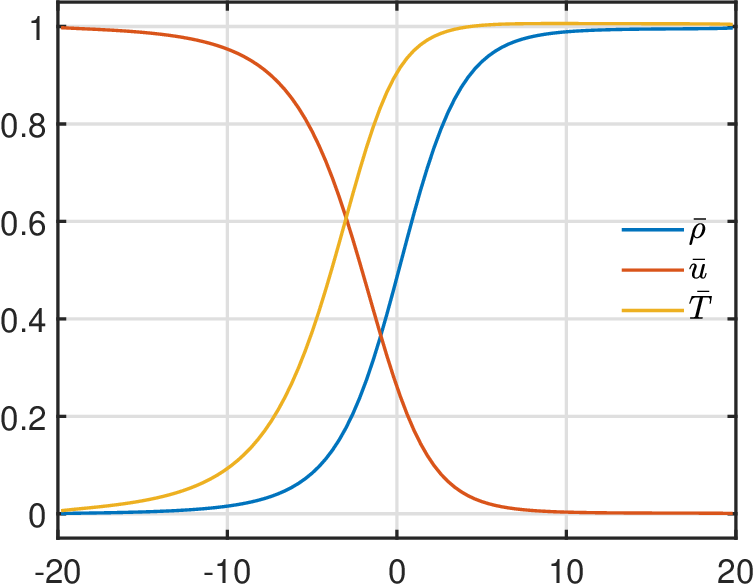}
        \label{fig:Ma3}
	 }	
      \subfigure[stress $\sigma$ and heat flux $q$]{
	 \includegraphics[width=0.45\linewidth, height=0.3\linewidth]{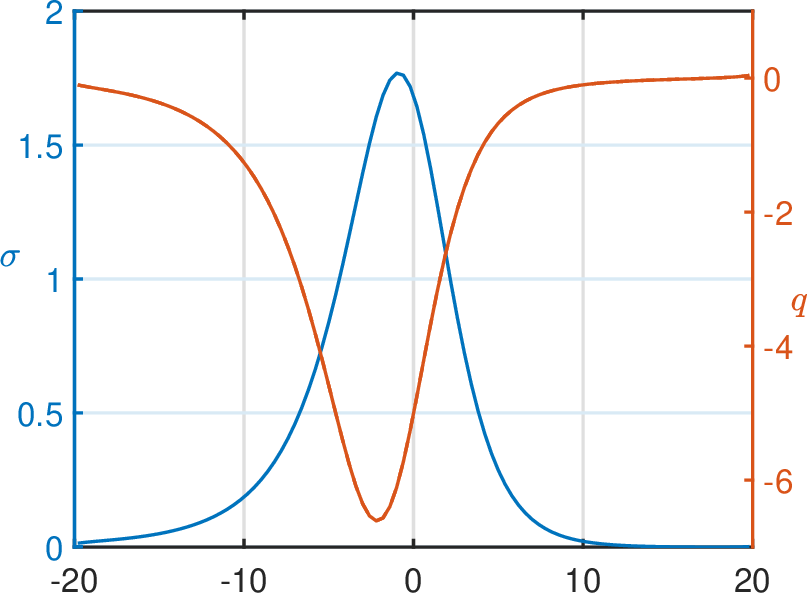}
        \label{fig:Ma3_sigma_q}
	 }
  \caption{Shock structures with Mach number $\Ma=3$.}
	\label{fig:Shock_Ma3}
\end{figure}

Three Mach numbers $\Ma = 1.55, 2$ and $3$ are considered in our tests. The results are given in Figs. \ref{fig:Shock_Ma1.55}, \ref{fig:Shock_Ma2} and \ref{fig:Shock_Ma3}. In these results, the density, velocity and temperature are normalized into the region $[0,1]$ by the following equations:
\begin{equation}
    \bar{\rho}=\frac{\rho-\rho_{\rm L}}{|\rho_{\rm R}-\rho_{\rm L}|},
    \qquad
    \bar{u}=\frac{u_1-u_{{\rm R},1}}{|u_{{\rm R},1}-u_{{\rm L},1}|},
    \qquad
    \bar{T}=\frac{T-T_{\rm L}}{|T_{\rm R}-T_{\rm L}|}.
\end{equation}
We have also plotted some nonequilibrium quantities including the stress and the heat flux defined by
\begin{displaymath}
  \sigma(x) = \int_{\mathbb{R}^3} |v_1 - u_1|^2 f(x,\bv) \,\mathrm{d}\bv - \rho(x) T(x), \qquad q(x) = \frac{1}{2} \int_{\mathbb{R}^3} |\bv - \bu(x)|^2 (v_1 - u_1) f(x,\bv) \,\mathrm{d}\bv.
  \label{eq:sigma_q}
\end{displaymath}
The general structures of shocks are correctly reflected in the numerical solutions: the temperature rises up earlier than the density in front of the shock wave; the thickness of the shock wave decreases with increasing Mach numbers for small Mach numbers; the nonequilibrium variables have larger magnitude for larger Mach numbers.

\begin{figure}[!htbp]
 \centering
      \subfigure[Decay of the relative difference between adjacent steps]{
      \label{fig:diff}
     \includegraphics[width=0.45\linewidth, height=0.3\linewidth]{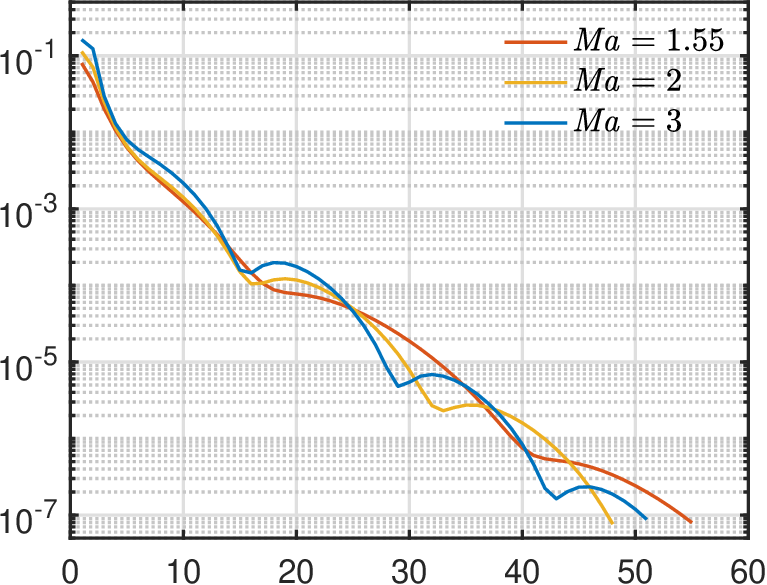}
     }
      \subfigure[Evolution of the residual $\|\bv \cdot \nabla_{\bx} f - \Kn^{-1} \mathcal{Q}(f,f)\|_2$]{
      \label{fig:residual}
     \includegraphics[width=0.45\linewidth, height=0.3\linewidth]{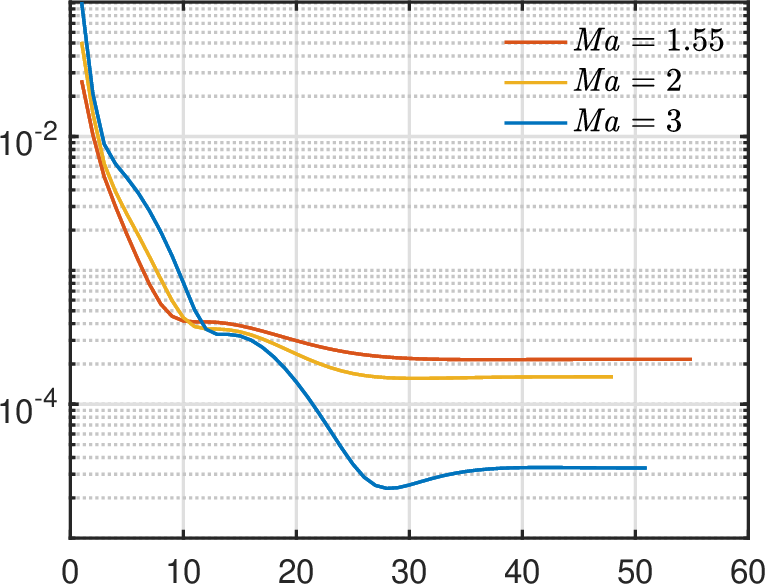}
     }
     \label{fig:iter_error2}
 	\caption{The evolution of the relative difference and the residual for shock structure simulations. In both figures, the $x$-axis is the number of iterations.}
	\label{fig:iteration}
\end{figure}

Fig. \ref{fig:iteration} demonstrates the convergence of our numerical method. 
Fig. \ref{fig:diff} shows the linear convergence rate for all Mach numbers, while Fig. \ref{fig:residual} shows that the residual plateaus after about 35 iterations. The reason that the residual does not decay to zero is likely to be the truncation of the spatial domain. Note that the final residual decreases as the Mach number increases, owing to the smaller shock thickness for larger Mach numbers.
The computational time is given in Tab. \ref{tab:computational_cost_shock_wave}. Since the number of grid points and the value of $\Kn$ are the same for all three Mach numbers, the computational time is generally proportional to the number of iterations.

\begin{table}[!ht]
\caption{Computational cost of shock structures.}
\begin{tabular}{c@{\qquad}c@{\qquad}c@{\qquad}c}
\hline
Mach number  & $\Ma=1.55$& $\Ma=2$ & $\Ma=3$          \\ \hline
Velocity domain        &  $[-7,7]^3$       &     $[-8,8]^3$       &  $[-11,11]^3$    \\
Computational time (s)   &     33752     &     29053   &     26911       \\  \hline
\end{tabular}
\label{tab:computational_cost_shock_wave}
\end{table}
 \section{Conclusion}
\label{sec:conclusion}
In this work, we tested the solver of the Boltzmann equation based on the symmetric Gauss-Seidel iteration, and the nonlinear system on each grid cell is solved using Newton's method (SGSN) and the fixed-point iteration (SGSFP). In our current tests, the SGSFP method is significantly faster than the SGSN method, but the SGSN method can achieve a stable computational cost in all regimes. Compared with the source iteration, the SGSFP method has faster convergence for small Knudsen numbers. However, unlike GSIS, the convergence rate of our method still depends on the Knudsen number, since our method does not utilize the macro-micro decomposition. Nevertheless, our approach does not require solving steady-state microscopic equations, and is therefore easier to implement. In general, we believe our approach is suitable for moderate to large Knudsen numbers ($\Kn \geqslant 0.05$). In our ongoing work, we are considering applying the micro-macro decomposition to the nonlinear algebraic equation  \eqref{eq:Iteration_F} to accelerate the inner iteration, which can possibly lead to an improved scheme that does not slow down for small Knudsen numbers.

\section*{Acknowledgements}
Tianai Yin was supported by the China Scholarship Council (CSC) (File No.202204890003). Zhenning Cai's work was supported by the Academic Research Fund of the Ministry of Education of Singapore under grant number A-0004592-00-00. This work of Yanli Wang is partially supported by the National Natural Science Foundation of China (Grant No. 12171026, U2230402, and 12031013), and the Foundation of the President of China Academy of Engineering Physics (YZJJZQ2022017). 

 \appendix
\section{Nondimensionalization of the Boltzmann equation} 
\label{sec:non-dimensionalization}
The nondimensionalization of the Boltzmann equation is based on the table below:
\medskip
\begin{center}
\begin{tabular}{cc@{\qquad}cc}\hline
\multicolumn{2}{c}{Original variable}             & Reference variable & Dimensionless variable             \\ \hline
\multicolumn{1}{c}{Mass}                  
& $m$      & $m_0$                        & $\tilde{m}= \frac{m}{m_0}$         \\
\multicolumn{1}{c}{Number density}        
& $n$   & $n_0$                   
& $\tilde{n}= \frac{n}{n_0}$         \\ 
\multicolumn{1}{c}{Temperature}           
& $T$   & $T_0$                        & $\tilde{T}= \frac{T}{T_0}$         \\ 
\multicolumn{1}{c}{Length}                
& $\bx$ & $L_0$                   
& $\tilde{\bx}= \frac{\bx}{L_0}$     \\ 
\multicolumn{1}{c}{Time}                  
& $t$   & $t_0=L_0/v_0$                   
& $\tilde{t}= \frac{t}{t_0} = \frac{t}{L_0/v_0}$         \\ 
\multicolumn{1}{c}{Velocity}              
& $\bv$ & $v_0=\sqrt{\frac{k}{m}T_0} $                      
& $\tilde{\bv}= \frac{\bv}{v_0} = \frac{\bv}{\sqrt{\frac{k}{m}T_0}}$ \\ 
\multicolumn{1}{c}{Distribution function} 
& $f$   &$f_0=n_0/v_0^3$                         & $\tilde{f}= \frac{f}{n_0/v_0^3}$   \\
\hline
\end{tabular}
\end{center}

\medskip

\noindent By such transformations, the Boltzmann equation can be reformulated as
\begin{displaymath}
\frac{\partial \tilde{f}}{\partial \tilde{t}} + \tilde{\bv} \cdot \nabla_{\tilde{\bx}} \tilde{f} = 
    \frac{L_0 n_0}{v_0} \int_{\bR^3} \int_{\bS^2}
    \mB(\bv-\bv_{\ast},\sigma)[\tilde{f}(\tilde{t},\tilde{\bx},\tilde{\bv}')\tilde{f}(\tilde{t},\tilde{\bx},\tilde{\bv}'_{\ast})-\tilde{f}(\tilde{t},\tilde{\bx},\tilde{\bv})\tilde{f}(\tilde{t},\tilde{\bx},\tilde{\bv}_{\ast})]
    \dd \sigma\dd \tilde{\bv}_{\ast}.
\end{displaymath}
Our desired form is
\begin{displaymath}
\frac{\partial \tilde{f}}{\partial \tilde{t}} + \tilde{\bv} \cdot \nabla_{\tilde{\bx}} \tilde{f} = 
    \frac{1}{\Kn} \int_{\bR^3} \int_{\bS^2}
    \tilde{\mB}(\tilde{\bv}-\tilde{\bv}_{\ast},\sigma)[\tilde{f}(\tilde{t},\tilde{\bx},\tilde{\bv}')\tilde{f}(\tilde{t},\tilde{\bx},\tilde{\bv}'_{\ast})-\tilde{f}(\tilde{t},\tilde{\bx},\tilde{\bv})\tilde{f}(\tilde{t},\tilde{\bx},\tilde{\bv}_{\ast})]
    \dd \sigma\dd \tilde{\bv}_{\ast}.
\end{displaymath}
By comparision, it can be easily seen that
\begin{equation} \label{eq:tilde_mB}
  \tilde{\mB}(\tilde{\bv}-\tilde{\bv}_{\ast},\sigma) =
  \frac{L_0 n_0}{v_0} \Kn \, \mB(\bv-\bv_{\ast},\sigma).
\end{equation}

For VHS gases, one can use \eqref{eq:chi_VHS} and \eqref{eq:mB} to derive that
\begin{equation} \label{eq:mB_new}
    \mB(\bv-\bv_{\ast},\sigma)=\frac{d^2}{4}\cdot|\bv-\bv_{\ast}|.
\end{equation}
The diameter of VHS gas molecules during the collision depends on the relative speed:
\begin{equation} \label{eq:d2}
    d^2 =d_0^2
    \left(\frac{4k T_0}{m |\bv - \bv_*|^2} \right)^{\omega-\frac{1}{2}}\frac{1}{\Gamma(2.5-\omega)},
\end{equation}
where $\omega$ is the viscosity index and $d_0$ is the reference molecular diameter at the reference temperature $T_0$ \cite{MolecularBird}. Recall that the Knudsen number $\Kn$ is defined as the ratio of the mean free path $\lambda_0$ to the characteristic length $L_0$, and the mean free path at the reference temperature is calculated by $\lambda_0 = 1/(\sqrt{2}\pi d_0^2 n_0)$. Therefore,
\begin{equation} \label{eq:Kn}
\Kn = \frac{1}{\sqrt{2}\pi d_0^2 n_0 L_0}.
\end{equation}
Now, we can plug \eqref{eq:mB_new}\eqref{eq:d2}\eqref{eq:Kn} into \eqref{eq:tilde_mB} to obtain
\begin{equation}
    \tilde{\mB}(\tilde{\bv}-\tilde{\bv}_{\ast},\sigma)=
    \frac{2^{2\omega-3}|\tilde{\bv}-\tilde{\bv}_{\ast}|^{2(1-\omega)}}{\sqrt{2}\pi\Gamma(2.5-\omega)}.
\end{equation}
For Maxwell molecules, $\omega = 1$, and thus $\tilde{\mB}(\tilde{\bv}-\tilde{\bv}_{\ast},\sigma)$ is a constant $1/(\pi\sqrt{2\pi})$.
\bibliographystyle{plain}
\bibliography{Reference}
\end{document}